\documentclass[%
reprint,
superscriptaddress,
frontmatterverbose, 
bibnotes,
amsmath,
amssymb,
aps,
floatfix,
]{revtex4-2}

\usepackage{graphicx}
\usepackage{dcolumn}
\usepackage{multirow}
\usepackage{bm}
\usepackage{amsmath}
\usepackage{amssymb}
\usepackage{amsfonts}

\usepackage{subfigure}
\usepackage{caption}
\usepackage{subcaption}
\usepackage{enumitem}
\usepackage{tabularx}
\usepackage{epstopdf}
\usepackage{ctable}
\usepackage{pbox}
\usepackage[normalem]{ulem}
\usepackage[justification=Justified]{caption}
\usepackage{color}
\usepackage{hyperref}
\hypersetup{
     colorlinks=true,
     linkcolor=blue,
     filecolor=magenta,      
     urlcolor=cyan,
     pdftitle={Overleaf Example},
     pdfpagemode=FullScreen,
     }

\newcommand{\SK}[1]{\textcolor{black}{{#1}}}

\begin{document}

\preprint{APS/123-QED}

\title{Effect of Random Pinning on the Yielding Transition of Amorphous Solid under Oscillatory Shear}

\author{Roni Chatterjee}
\author{Monoj Adhikari}
\author{Smarajit Karmakar}%
\affiliation{%
 Tata Institute of Fundamental Research, 36/P, Gopanpally Village, Serilingampally Mandal, Ranga Reddy District, Hyderabad 500046, Telangana, India\\
}%

\date{\today}

\begin{abstract}
\SK{We investigate the effects of random pinning, where we freeze the relaxation degrees of freedom for a fraction of randomly selected particles, on the yielding transition under oscillatory shear through extensive computer simulations. Using Kob-Anderson model as our model glass former, we pin a fraction of the particles. These pinned particles can move affinely under the imposed oscillatory shear deformation but are not allowed to relax through subsequent rearrangements due to plastic events. As the fraction of pinned particles increases, the system transitions from being a fragile glass former to a strong glass former. This gives us opportunity to examine how changes in fragility impact the yielding transition. Our results demonstrate notable differences in the yielding transition between strong and fragile glass formers under random pinning. This aligns with previous observations where fragility was altered by changing the packing fraction in soft sphere model albeit with a main difference that random pinning significantly suppresses the formation of shear bands, even in well-annealed glass samples.} 
\end{abstract}

\maketitle


\noindent{\bf Introduction:}
\SK{The mechanical behavior of amorphous solids is critical due to their wide-ranging applications, from bulk systems to nanoscale technologies, which impact our daily lives \cite{fielding2000aging,falk2011deformation,schuh2007mechanical,bonn2017yield,cochran2022slow,Kbhattacharya,naturePine,natphysCorte,PhysRevLett.107.010603,PhysRevE.88.032306,PhysRevLettPaulsen}. In recent years, the response of amorphous solids to external deformation has been extensively studied \cite{argon1979plastic,princen1985rheology,schall2007structural,varnik2004study,tsamados2009local,irani2014impact,karmakar2010plasticity,ozawa2020role,liu2022fate,dauchot2011athermal}. Despite lacking long-range order, amorphous materials exhibit a response similar to that of crystalline solids under small deformation amplitudes. However, at larger amplitudes, significant plastic deformation occurs, eventually leading to a flowing state at what is known as the yielding amplitude. Studies have shown that the nature of the yielding transition is highly sensitive to the preparation history or age of the sample \cite{yeh2020glass,bhaumik2021role,ozawa2018random,barlow2020ductile,ozawa2023creating,singh2020brittle,liu2021oscillatory,bhaumik2022avalanches,MemFiocco}. Glasses prepared from high-temperature melts (poorly annealed) behave differently from those prepared at low-temperature melts (well annealed), both under uniform and oscillatory shear.}

\SK{Under cyclic shear deformation, amorphous solids exhibit an absorbing phase transition near the yielding point. Below the yielding amplitude, the system reaches an absorbing state where stroboscopic configurations remain unchanged after a transient \cite{krishnan2023annealing,fiocco2013oscillatory,leishangthem2017yielding,JPCMFiocco,barbot2020rejuvenation,bandi2018training,adhikari2018memory,keim2019memory}. Above yielding, the system becomes diffusive, with configurations continually changing. While this transition from absorbing to diffusive states occurs in glasses with different annealing histories, the nature of the transition depends on the degree of annealing. In poorly annealed glasses, stroboscopic energy decreases with increasing deformation until the yielding amplitude, after which it rises. Glasses prepared above a threshold energy exhibit the same behavior, yielding at the same amplitude. Conversely, well-annealed glasses, prepared below a threshold energy corresponding to the Mode Coupling Temperature ($T_{MCT}$), maintain constant stroboscopic energy until the yielding amplitude is reached, followed by an energy increase. As the degree of annealing increases, the yielding amplitude shifts to larger values. 
Similar to energy, the stress-strain relationship shows distinct behaviors depending on the annealing history of the samples. Poorly annealed glasses exhibit gradual, ductile yielding, while well-annealed glasses display sudden, brittle failure. This change in yielding behavior with the degree of annealing is consistent under both uniform and oscillatory shear deformation. However, whether this transition persists in the thermodynamic limit remains an open question.}

\SK{Recent work on assemblies of soft spheres interacting via harmonic potentials demonstrates that the fragility of the initial glass former plays a key role in determining the nature of the yielding transition under cyclic shear \cite{chatterjee2024role}. Fragility refers to how rapidly viscosity or relaxation time increases as temperature decreases in the supercooled liquid state. By adjusting density and thus fragility, a strong link between the kinetic properties of liquids and the mechanical behavior of amorphous solids has been observed. Both strong and fragile glass formers exhibit similar yielding responses in poorly annealed states, but with better annealing, a clear distinction emerges. Fragile glass formers show a significant increase in yielding strain with improved annealing, whereas strong glass formers exhibit minimal change. This contrast is also reflected in their stress-strain curves: strong glass formers consistently show ductile-like yielding, while fragile glass formers become increasingly brittle, characterized by a large stress jump at the yield point with increasing annealing.}

\SK{Previous studies have shown that fragility can also be changed in systems where a fraction of particles are randomly pinned, frozen in time \cite{cammarota2012ideal,chakrabarty2015dynamics,karmakar2013random,ozawa2015equilibrium,brito2013jamming,fullerton2014investigating}. By varying the fraction of pinned particles, the system’s fragility can be tuned. These systems exhibit interesting yielding behavior under uniform shear \cite{bhowmik2019effect}, showing homogeneous yielding without the formation of shear bands, which are typically observed in unpinned systems at poorly annealed states \cite{shi2006atomic,parmar2019strain,hassani2019probing,shrivastav2016yielding,Maloney2021,keim2022mechanical,urbani2017shear,lamp2022brittle}. We now ask: what happens when these systems are subjected to oscillatory shear? Does changing the pinning fraction, and thereby the fragility, influence yielding behavior in a manner similar to previous studies? Does yielding remain homogeneous even under cyclic deformation? Additionally, we have studied these systems in two dimensions to examine how dimensionality impacts their yielding behavior.}
\begin{figure*}[htp]
   \includegraphics[width=0.45\textwidth]{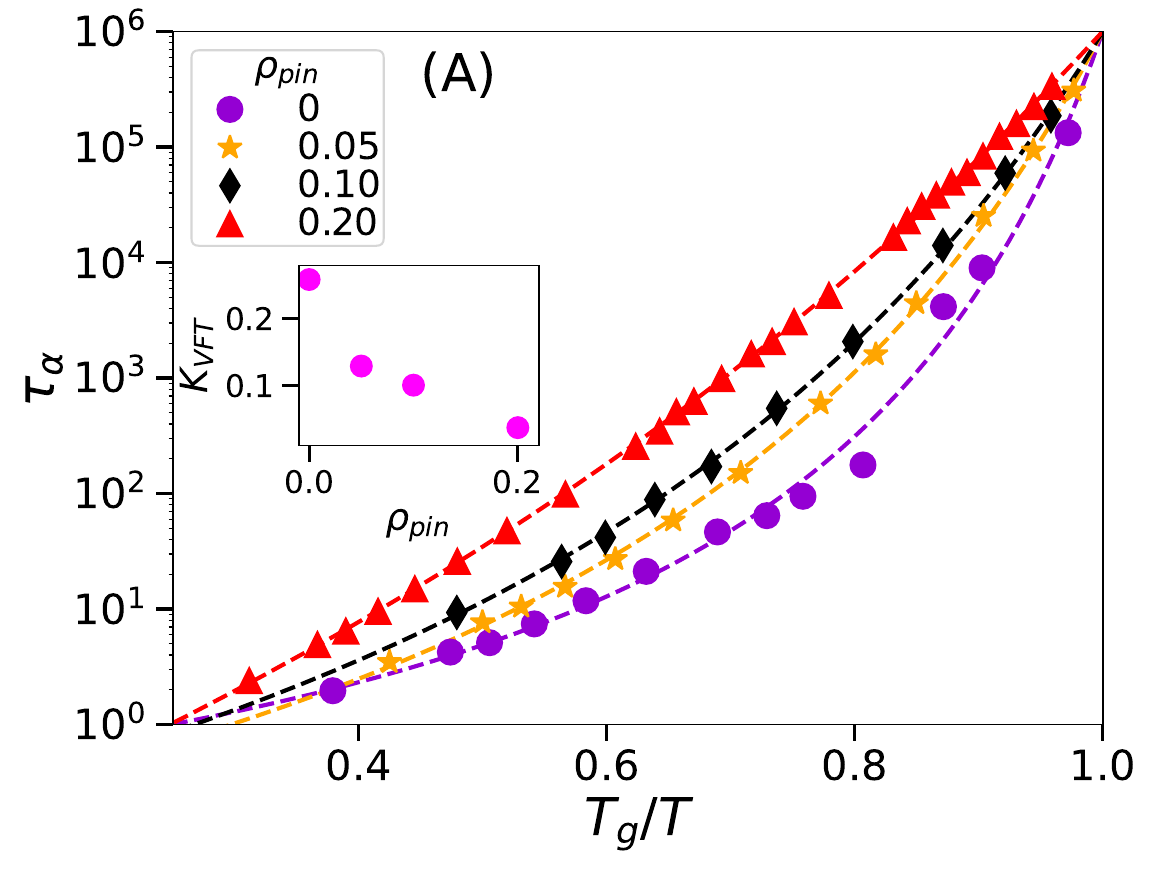}
   \includegraphics[width=0.45\textwidth]{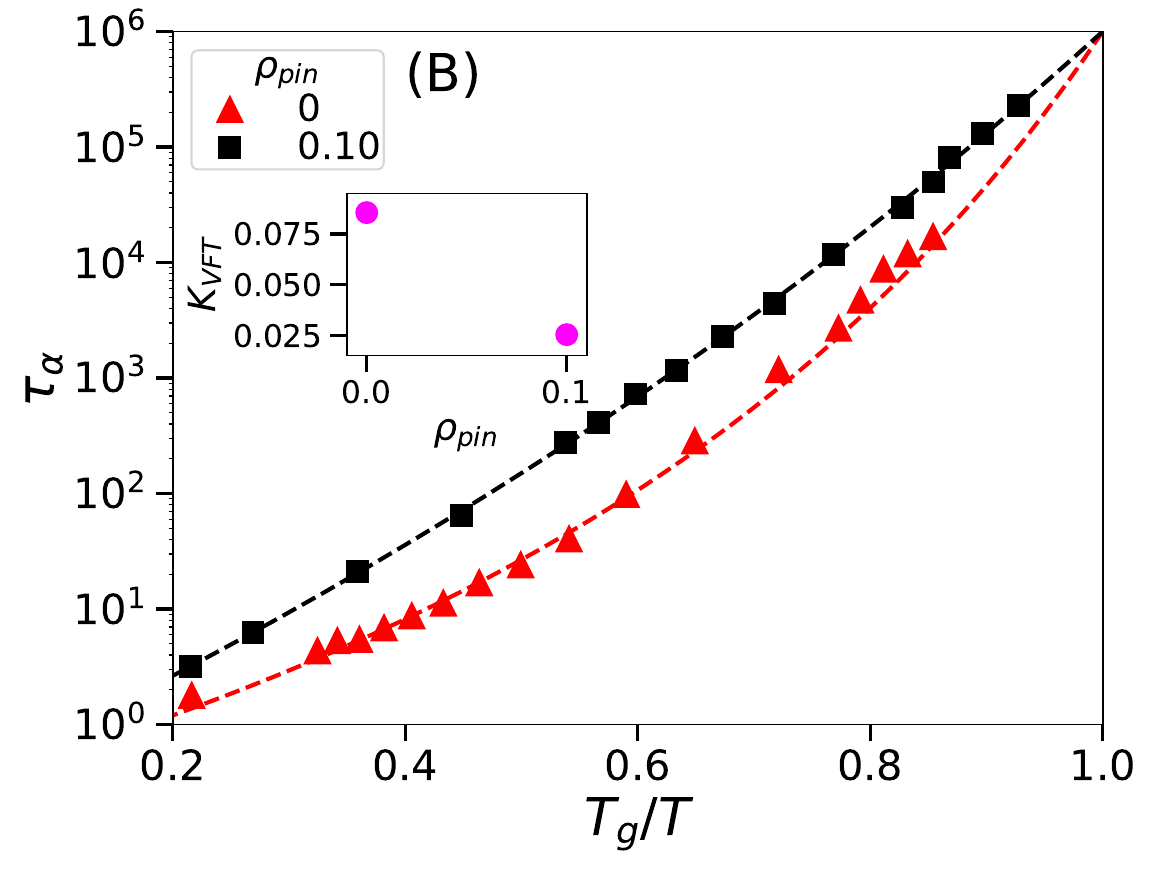}
  \caption{Angell plot: Relaxation time ($\tau_\alpha$) is plotted against rescaled temperature $T_g/T$ where $T_g$ is the calorimetric glass transition temperature (see text). A curve in this plot showing linear behaviour will represent glasses belonging to strong glass-formers. Panel (A) and (B) show results for the three-dimensional (3D) and two-dimensional (2D) Kob-Anderson model, a typical fragile glass-former at zero pinning.}
\label{fig:1}
\end{figure*}

\begin{figure*}[htp]
  \centering
   \includegraphics[width=0.325\textwidth]{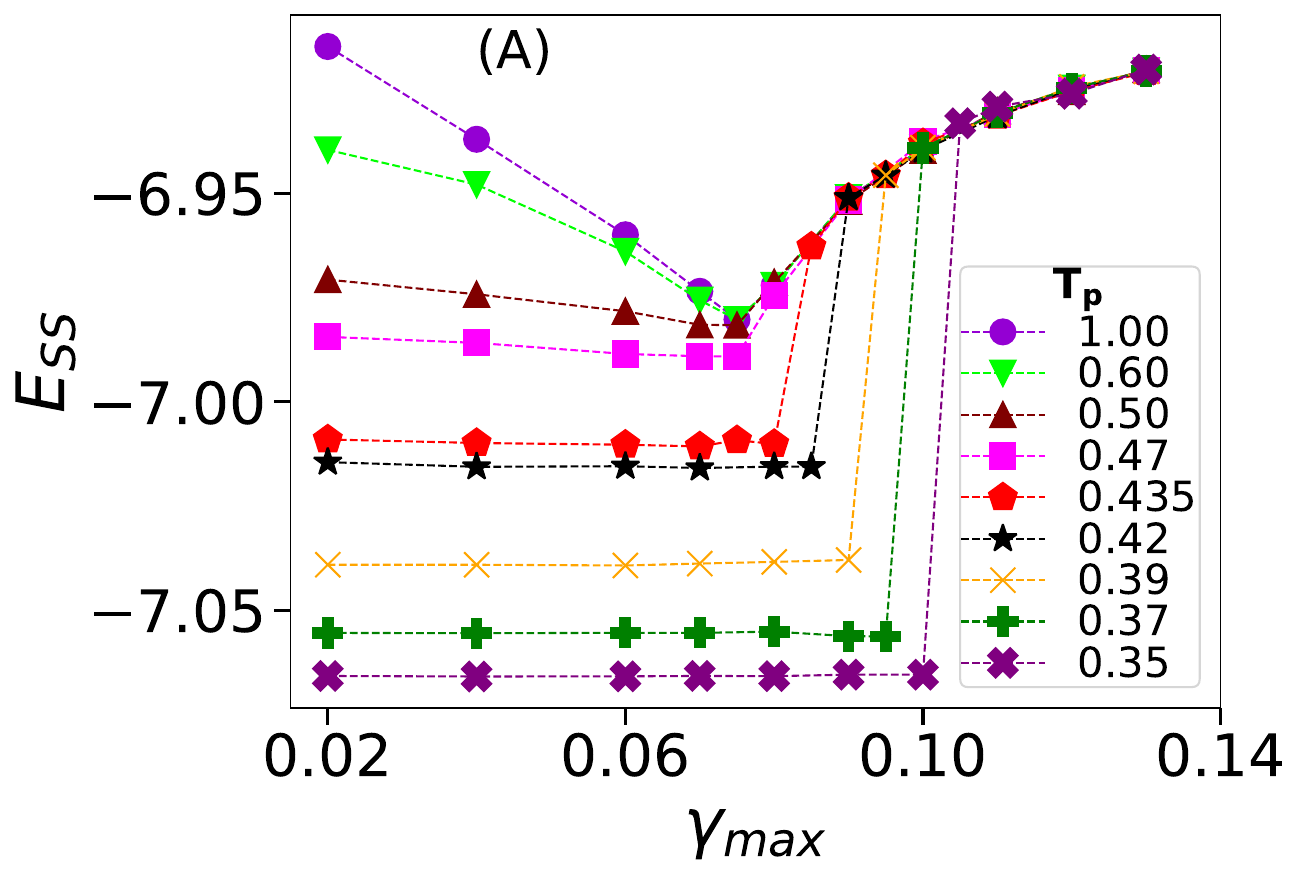}
   \includegraphics[width=0.325\textwidth]{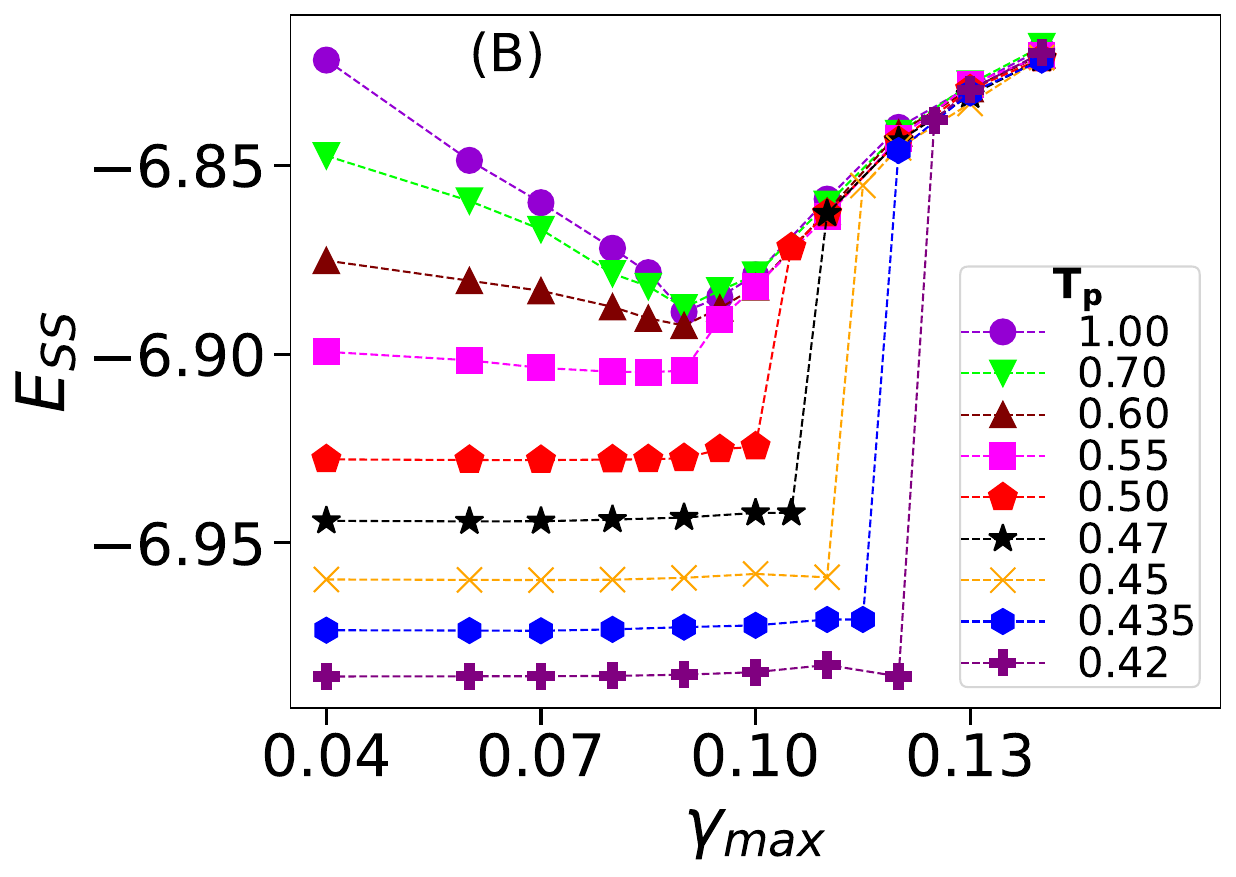}
   \includegraphics[width=0.325\textwidth]{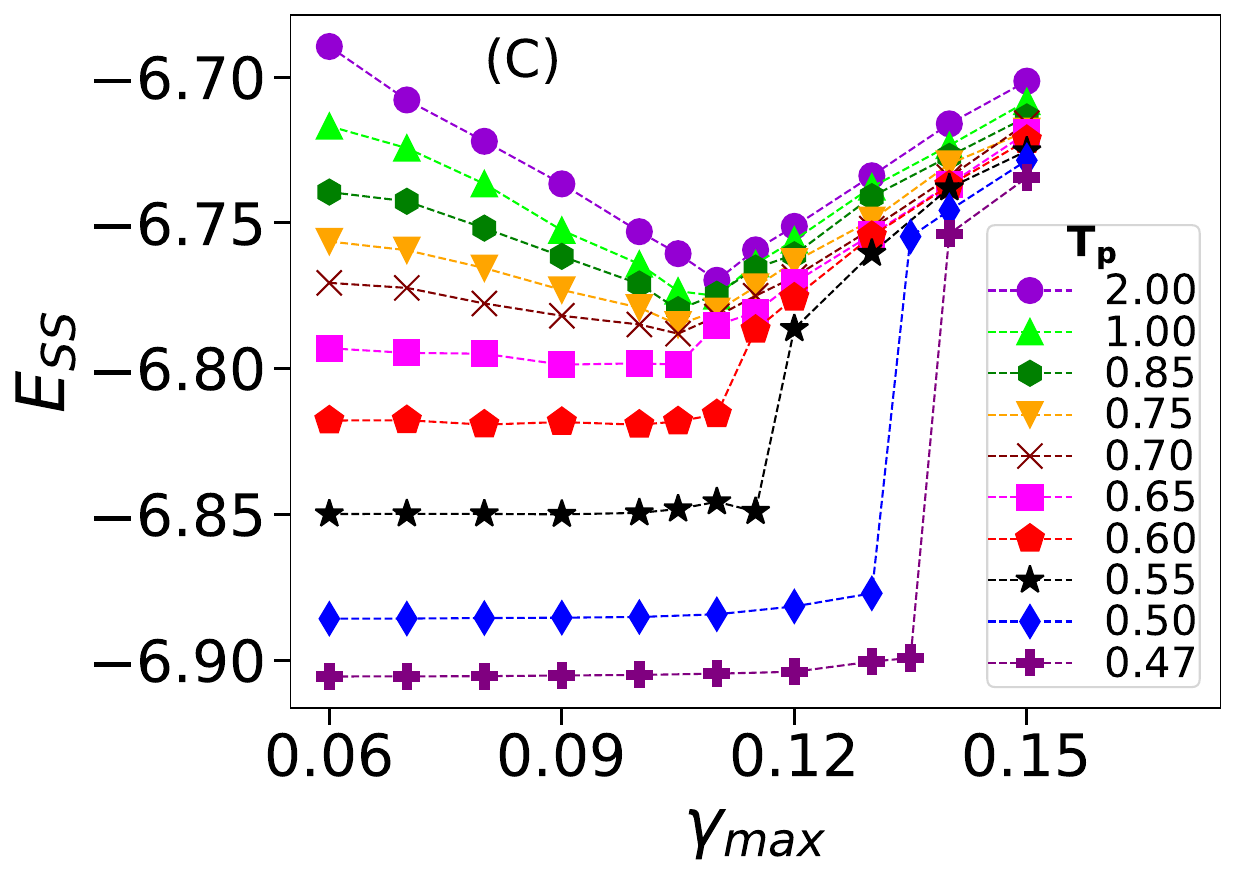}
   \includegraphics[width=0.325\textwidth]{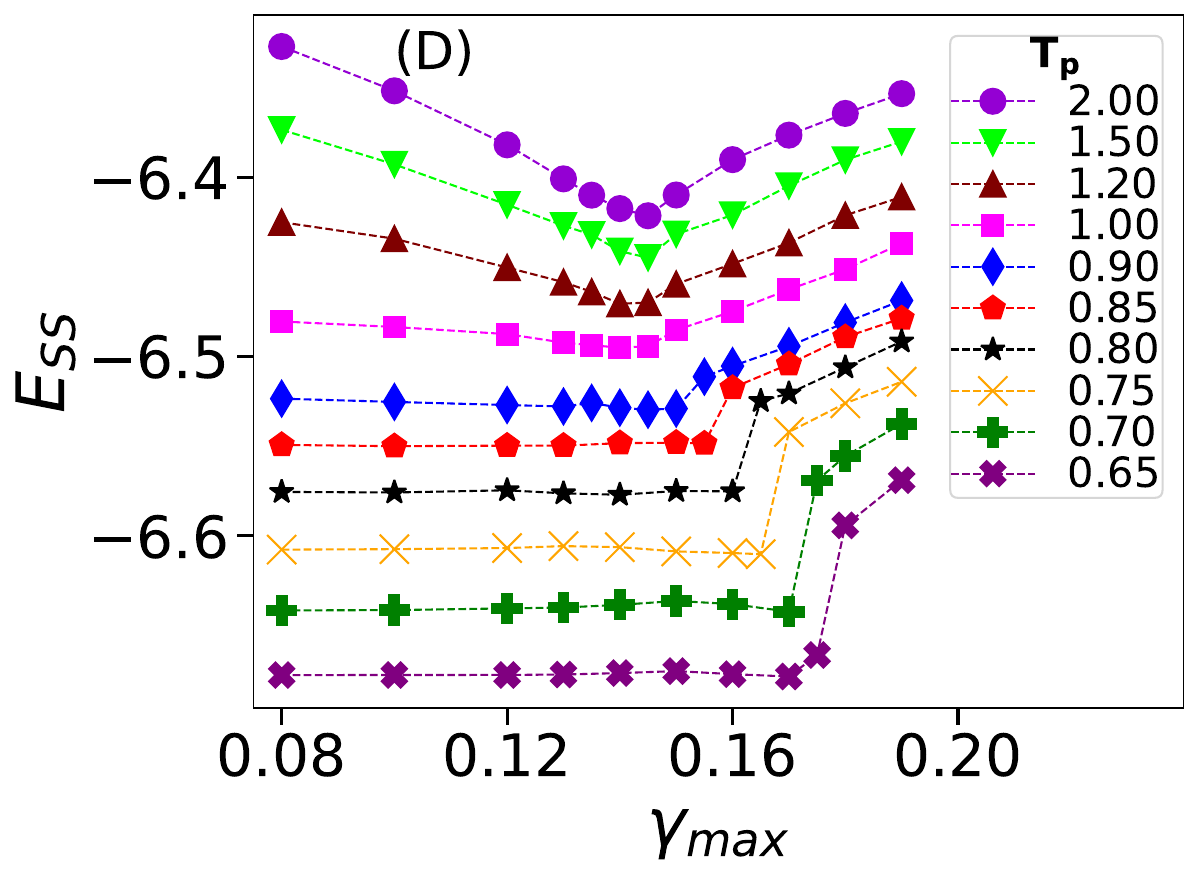}
   \includegraphics[width=0.325\textwidth]{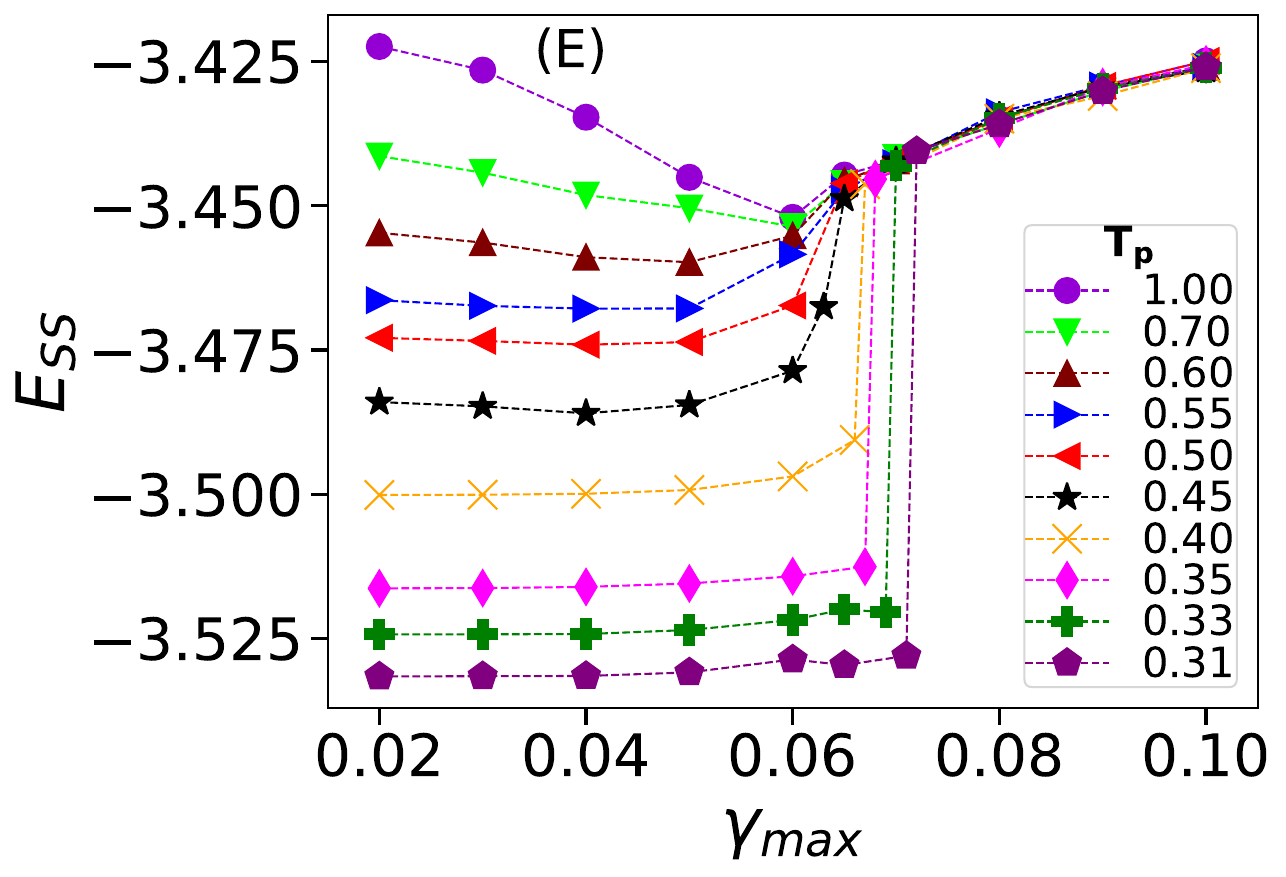}
   \includegraphics[width=0.325\textwidth]{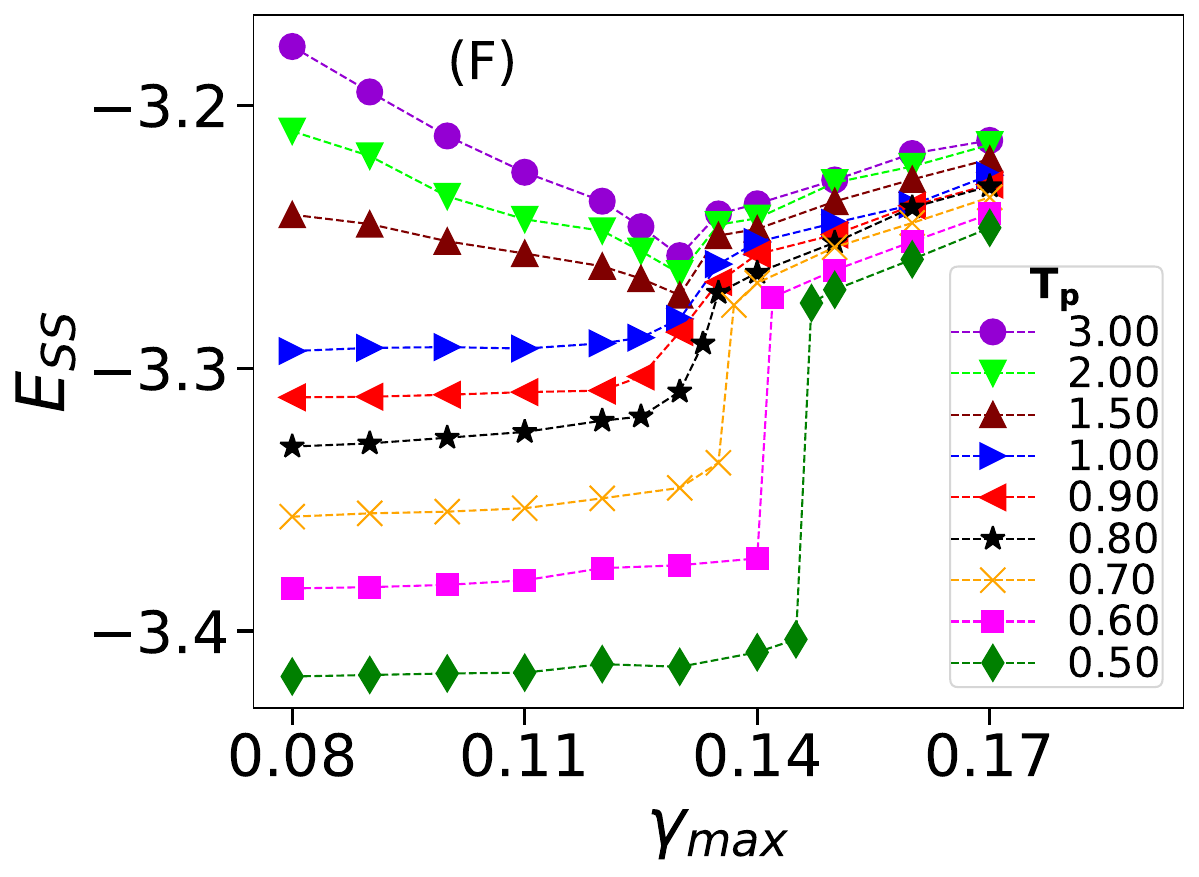}
  \caption{Steady state energy ($E_{SS}$) is plotted against cyclic shear amplitude $\gamma_{max}$ for unpinned ($\rho_{pin} = 0$) system in (A). All the poorly annealed samples ($T_p > 0.435$) show mechanical annealing till their common yield point at $\gamma_{max} = \gamma_c \approx 0.075$. The well annealed samples ($T_p \leq 0.435$) shows no mechanical annealing till yielding ($\gamma_{max} < \gamma_y$) but after yielding there is a sharp discontinuous energy jump. With increasing annealing below $T_p = 0.435$, the yield point $\gamma_y$ gradually increases. $T_p = 0.435$ denotes the $T_{MCT}$ for $\rho_{pin} = 0$ system. $E_{SS}$ vs $\gamma_{max}$ for different $T_p$ is plotted in the pinned system - (B) $\rho_{pin} = 0.05$, (C) $\rho_{pin} = 0.10$, (D) $\rho_{pin} = 0.20$ in 3D. The striking difference from the unpinned system is that after yielding difference in steady state energy increases with pin concentration $\rho_{pin}$. The same is also plotted for (E) $\rho_{pin} = 0$ and (F) $\rho_{pin} = 0.10$ in 2D.}
\label{fig:2}
\end{figure*}

\vskip +0.1in
\noindent{\bf Simulation details: }
\SK{We have used a binary mixture of Kob-Anderson model glass in three dimensions (3D) \cite{kob1994scaling,kob1995testing,kob1995testing} and two dimensions (2D) \cite{bruning2008glass}. The mixing ratio of A and B-type particles is $80:20$ and $65:35$ in 3D and 2D, respectively. The interaction potential is given by
\begin{equation*}
V_{\alpha\beta}(r)=4\epsilon_{\alpha\beta}\left[\left(\frac{\sigma_{\alpha\beta}}{r}\right)^{12}-\left(\frac{\sigma_{\alpha\beta}}{r}\right)^6+C_0+C_2\left(\frac{r}{\sigma_{\alpha\beta}}\right)^2\right]
\end{equation*}
where $\alpha,\beta \in {A,B}$; interaction strength of pairs $\epsilon_{AB}/\epsilon_{AA} = 1.5, \epsilon_{BB}/\epsilon_{AA} = 0.5$; and diameters $\sigma_{AB}/\sigma_{AA} = 0.8, \sigma_{BB}/\sigma_{AA} = 0.88$. The interaction cut-off is at $r_c = 2.5 \sigma_{\alpha\beta}$ and number density $\rho = 1.2$. We chose the number of particles $N = 5000$ in 3D and $N = 2000$ in 2D. We have performed NVT (constant particle number N, volume V, temperature T) molecular dynamics simulation using the Nose-Hover thermostat in a cubic box in 3D and a square box in 2D, maintaining periodic boundary conditions. Integration time step $dt = 0.005$ is used. We equilibrated the liquids over a wide range of temperatures where relaxation time $\tau_\alpha$ reaches up to $10^6$ in both dimensions. For each temperature, we simulated $12$ samples in 3D and $16$ samples in 2D. To study the pinning effect on yielding transition, we randomly pinned a fraction of particles in the equilibrated configuration. We studied pin concentration $\rho_{pin}$ $0$,$5\%$,$10\%$,$20\%$ in 3D and $0$, $10\%$ in 2D.}

\SK{After equilibration of the liquids at their parent temperatures ($T_p$), we performed the conjugate gradient method (CG) to get the minimized configurations. These states are also called inherent structures (IS). We applied cyclic shear on these IS configurations for different strain amplitude ($\gamma_{max}$). The strain sequence $0 \rightarrow \gamma_{max} \rightarrow -\gamma_{max} \rightarrow 0$ is defined as one single cycle. In 3D we sheared the system in x-z direction via athermal quasistatic (AQS) shear protocol \cite{maloney2006amorphous} where (I) each atom is moved by an affine displacement ($r^\prime_x \rightarrow r_x + r_z \times d\gamma, r^\prime_y \rightarrow r_y, r^\prime_z \rightarrow r_z$) followed by (II) energy minimisation via CG. The chosen elementary strain is $d\gamma = 2 \times 10^{-4}$. In 2D, we sheared the system in the x-y direction via the same protocol. During shear we applied Lees-Edwards boundary condition\cite{LeesEdwards}.}

\SK{While shearing the pinned system, all the particles (including pinned particles) undergo affine displacement, but during minimization, the pinned particles remain at their respective affinely displaced position. We make this happen by setting each force component to zero on the pinned particles while performing CG minimization. Due to this protocol, some cases might appear while shearing at various $\gamma_{max}$ that pinned particles come very close to each other, and the corresponding energy of the system might blow up. To get rid of the overlapping of pinned particles during shear at different $\gamma_{max}$, we choose pinned particles in such a way that at least they are separated by distance $2^{1/6}\sigma$ during the application of $\gamma_{max}$.}

\vskip +0.1in
\noindent{\bf Results: }
\begin{figure*}[!htpb]
  \centering
   \includegraphics[width=0.45\textwidth]{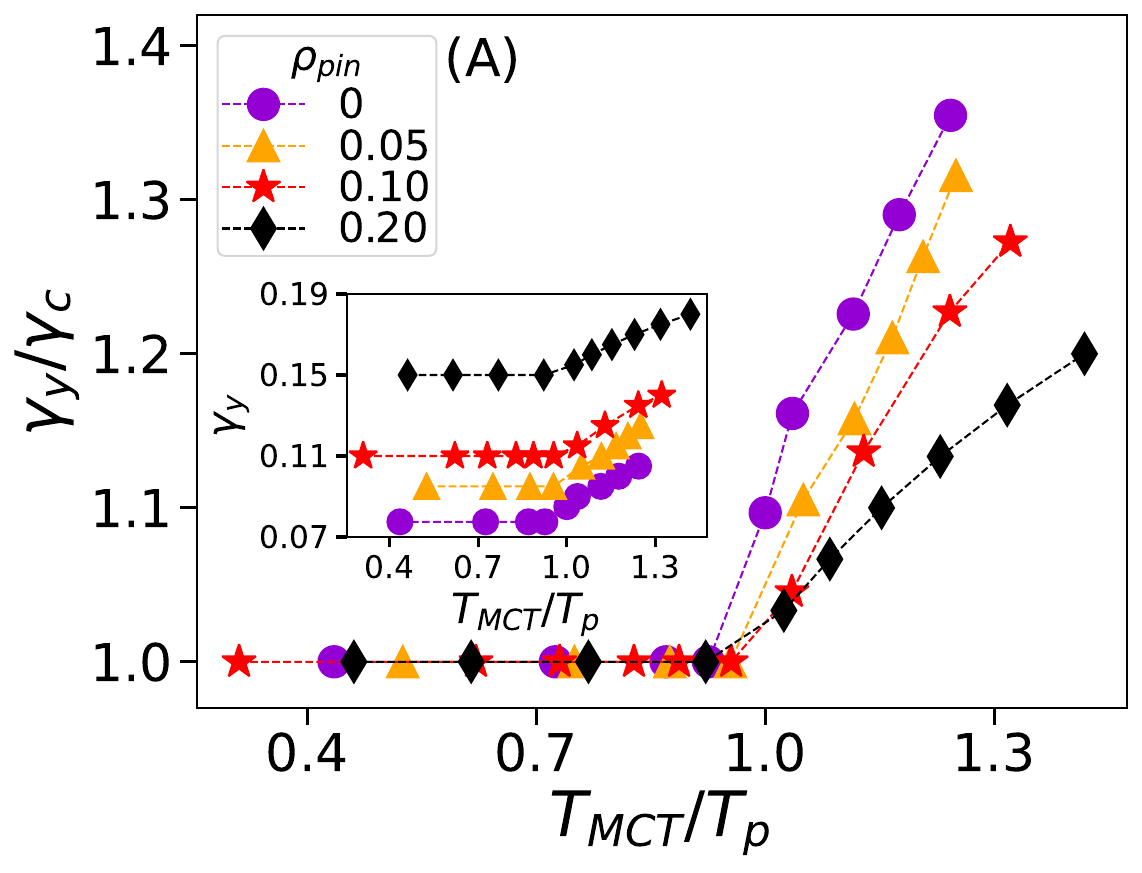}
   \includegraphics[width=0.45\textwidth]{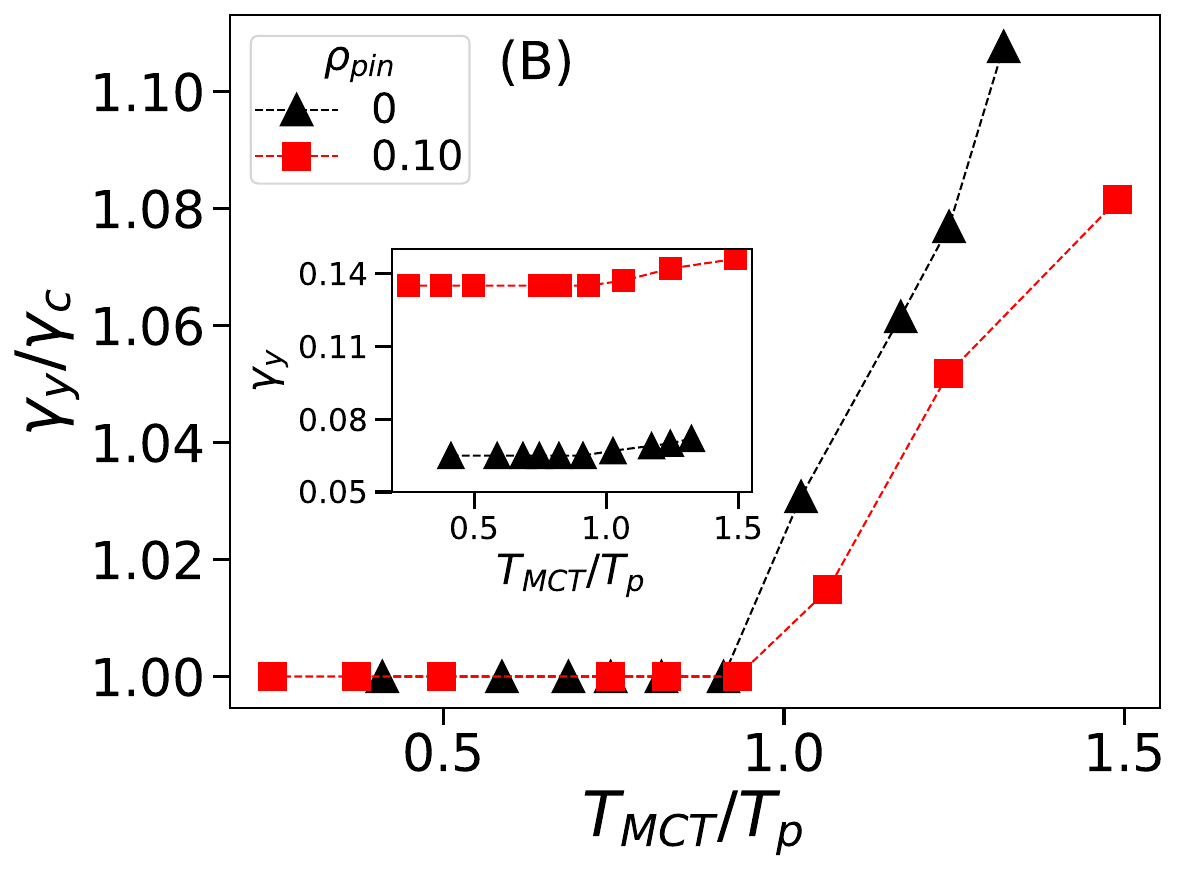}
  \caption{Shifting of yield point ($\gamma_y$) with respect to common yield point ($\gamma_c$) is shown with annealing in (A) 3D and (B) 2D. Temperatures are normalised by $T_{MCT}$ (from where $\gamma_y$ starts to shift from $\gamma_c$). Inset show the variation of $\gamma_y$ with annealing in both dimensions.}
\label{fig:3}
\end{figure*}
\SK{In Fig.\ref{fig:1}, we show the relaxation time, $\tau_{\alpha}$, as a function of temperature scaled with the kinetic glass transition temperature $T_g$ (at which $\tau_\alpha = 10^6$) for different pinning fractions, $\rho_{pin}$, in two dimensions: 3D (left panel) and 2D (right panel). We computed $\tau_{\alpha}$ as the time at which the overlap function, $q(t=\tau_\alpha) = 1/e$. In the inset of each panel, we present the kinetic fragility, $K_{VFT}$, obtained by fitting $\tau_{\alpha}$ vs. $T_p$ data with the Vogel-Fulcher-Tammann (VFT) \cite{stickel1995dynamics} formula: 
$\tau_{\alpha} = \tau_0 \exp\left(\frac{1}{K_{VFT}\left(\frac{T}{T_{VFT}} - 1\right)}\right)$,
where $K_{VFT}$ is the kinetic fragility, and $T_{VFT}$ is the temperature at which $\tau_{\alpha}$ diverges. Increasing the pinning concentration, $\rho_{pin}$, decreases the fragility of the liquid, making it stronger in both dimensions. In 3D, fragility decreases up to 7 times, from $ \rho_{pin} = 0 $ $( K_{VFT} = 0.258$) to $\rho_{pin} = 0.20$ $( K_{VFT} = 0.036$). Similarly, in 2D, fragility decreases by a factor of 3, from $ \rho_{pin} = 0$, $( K_{VFT} = 0.074$) to $\rho_{pin} = 0.10$ $( K_{VFT} = 0.025 $). While studying the oscillatory shear response of the system, the temperatures shown in the legend correspond to the temperature at which the liquid was initially equilibrated, and we refer to it as the parent temperature $T_p$ in the rest of the article.}


\begin{figure*}[htpb]
  \centering
   \includegraphics[width=0.325\textwidth]{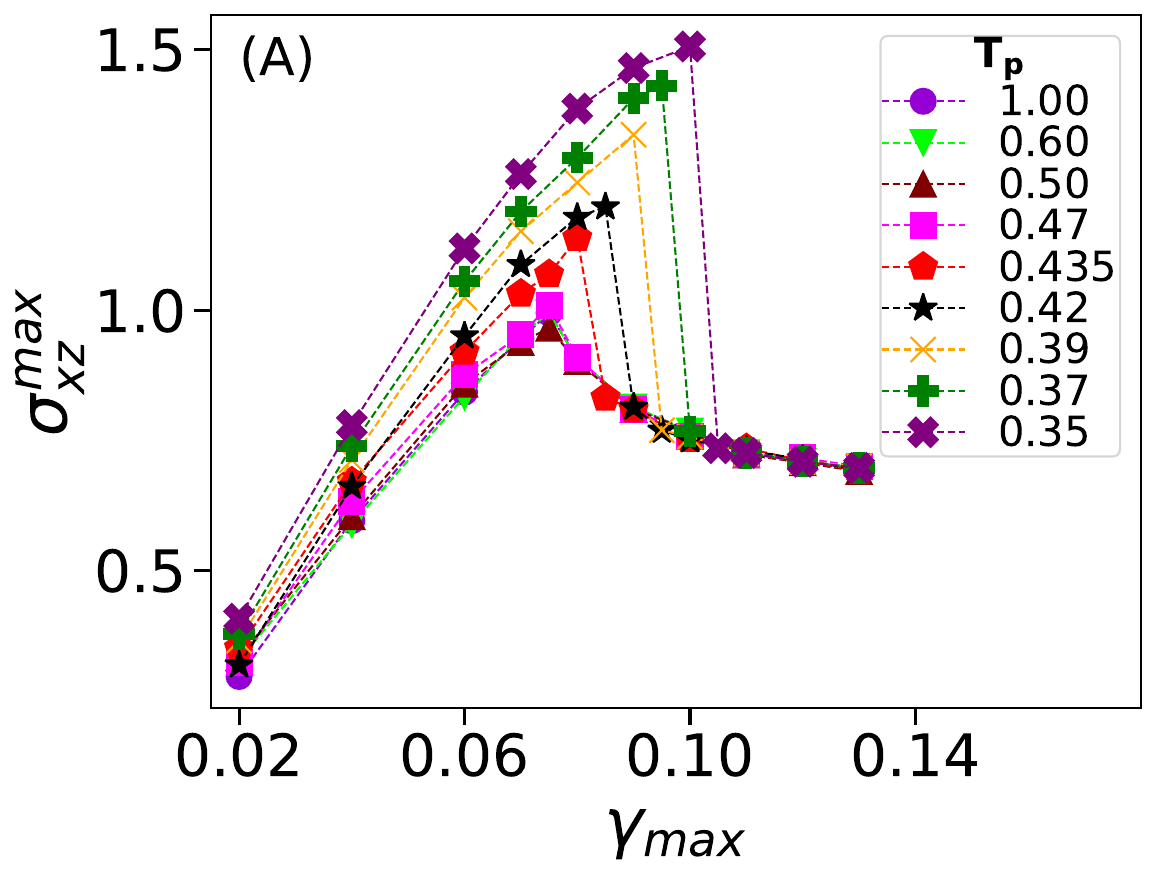}
   \includegraphics[width=0.325\textwidth]{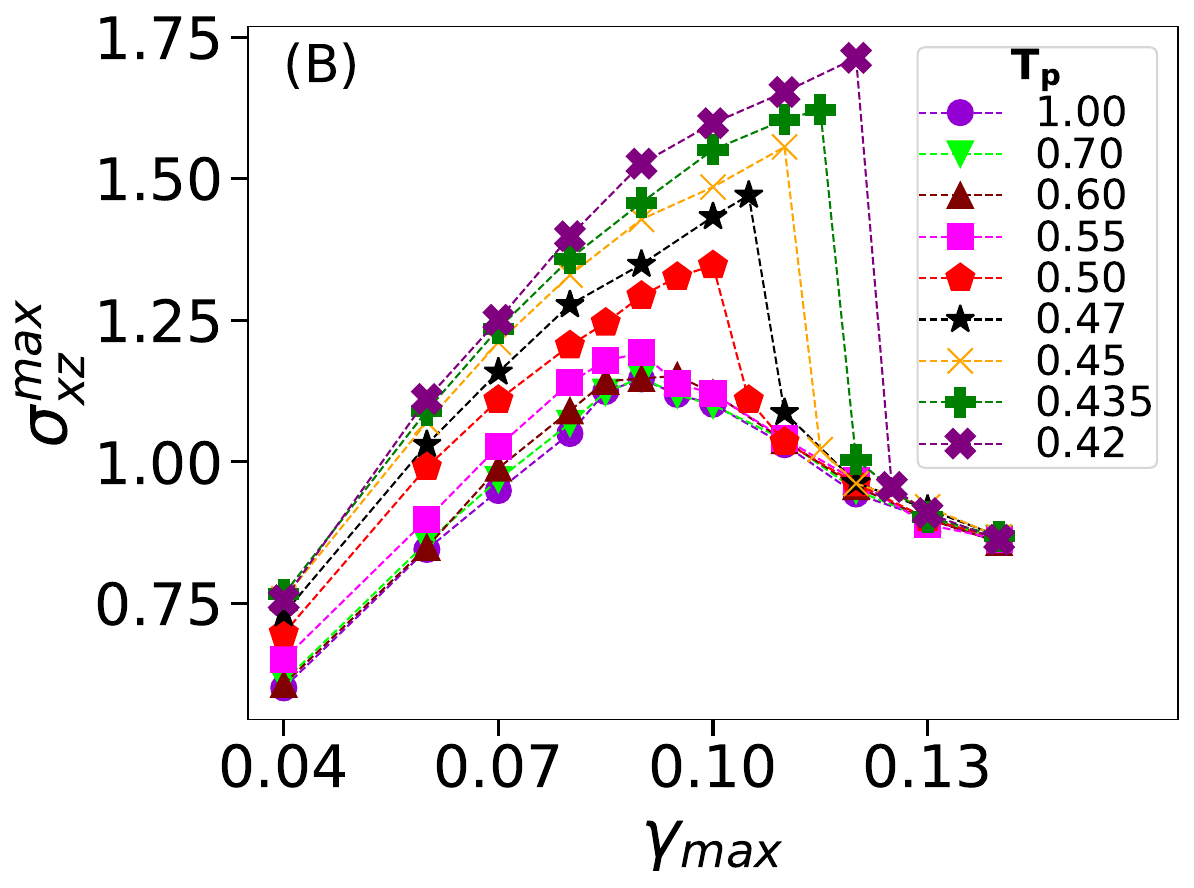}
   \includegraphics[width=0.325\textwidth]{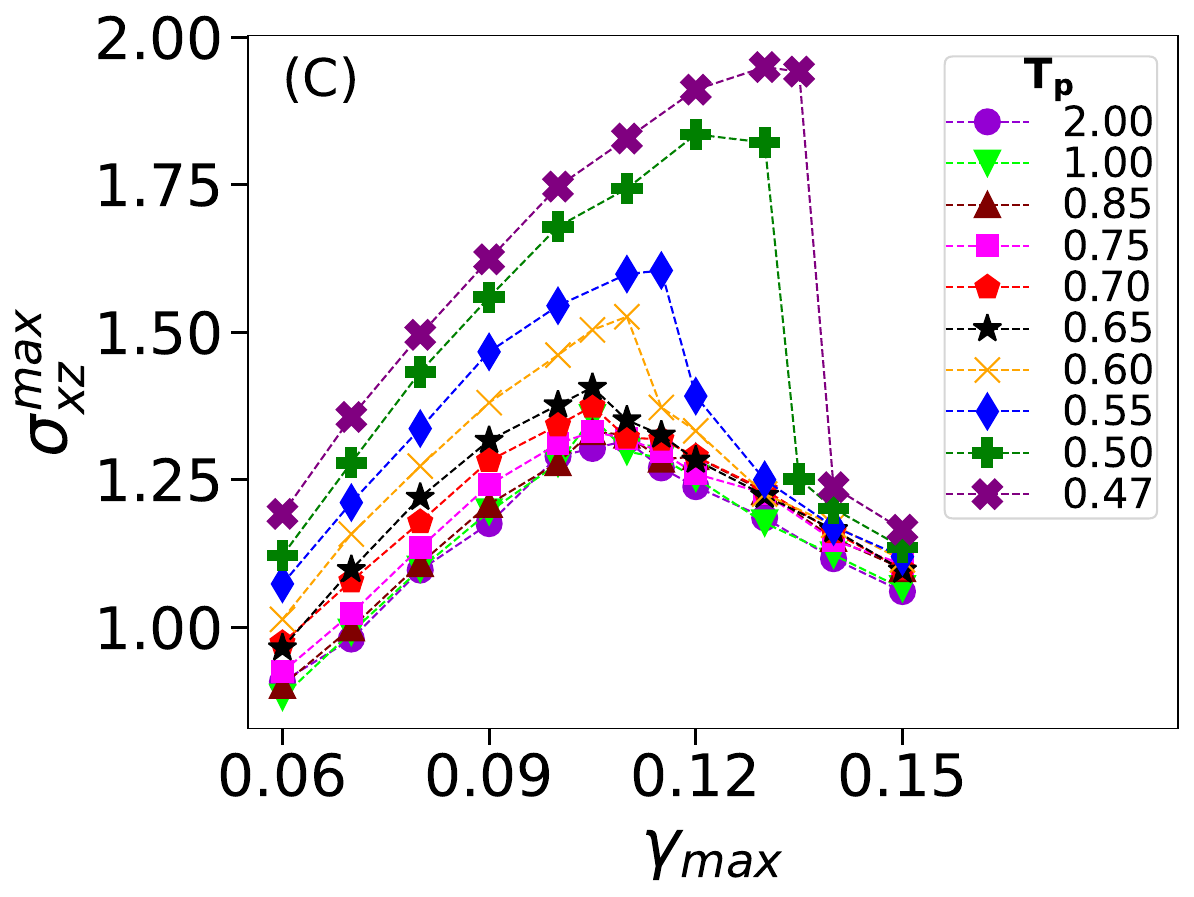}
   \includegraphics[width=0.325\textwidth]{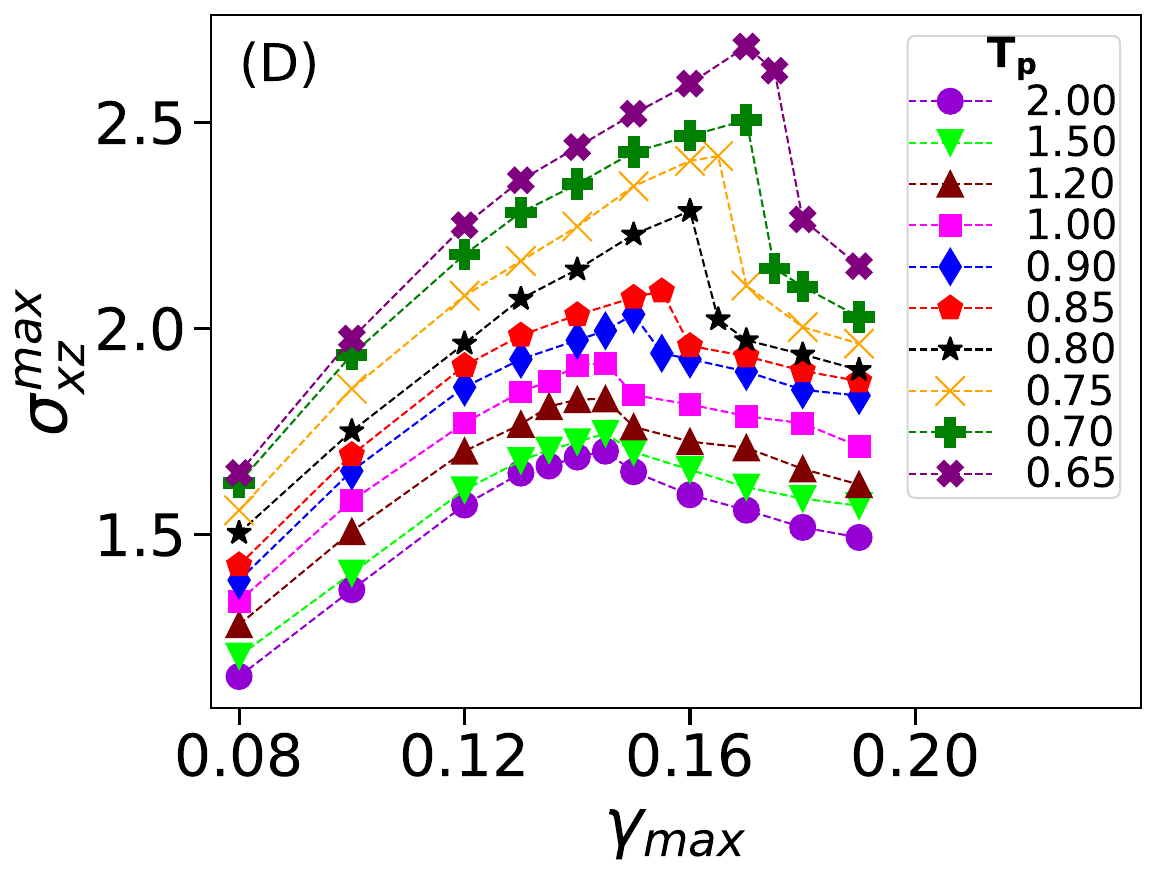}
   \includegraphics[width=0.325\textwidth]{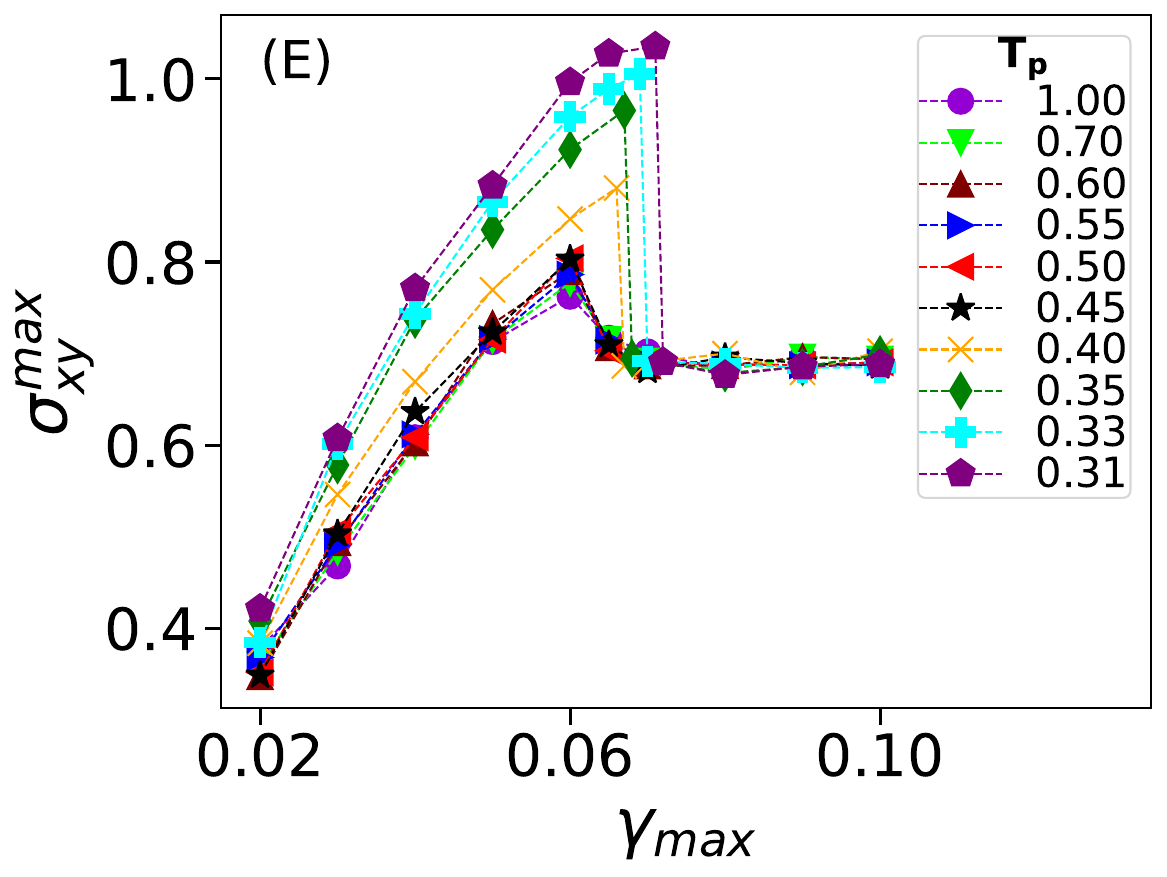}
   \includegraphics[width=0.325\textwidth]{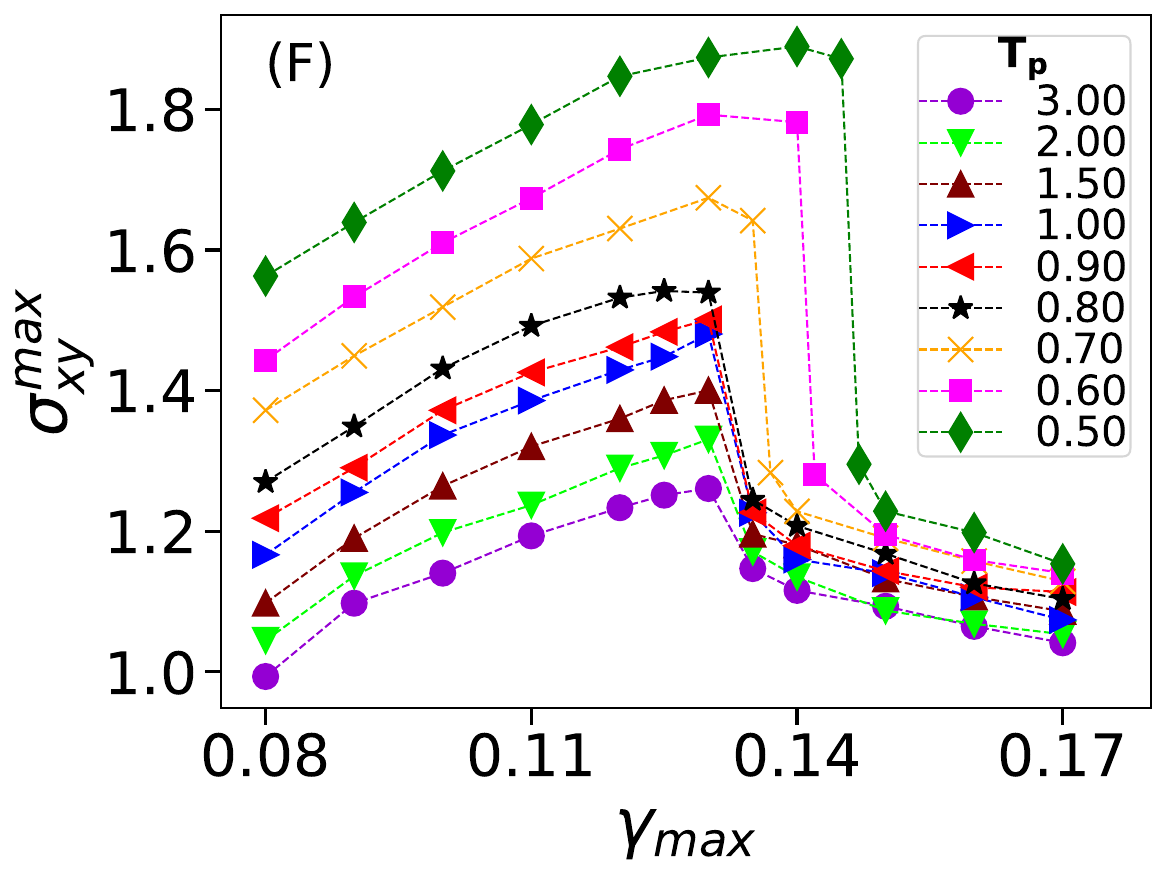}
  \caption{Steady state maximum stress ($\sigma^{max}_{xz}$) is plotted against cyclic shear amplitude $\gamma_{max}$ for unpinned ($\rho_{pin} = 0$) system in (A). In the case of the well-annealed samples ($T_p \leq 0.435$), we see there is a sharp stress drop after yielding, indicating brittle characteristics. For the pinned systems - (B) $\rho_{pin} = 0.05$, (C) $\rho_{pin} = 0.10$, (D) $\rho_{pin} = 0.20$ in 3D, this stress drop decreases indicating more ductile features. Here also, after yielding the difference in steady state stress $\sigma^{max}_{xz}$ increases with pin concentration $\rho_{pin}$. Similar plots are also shown in case of 2D for (E) $\rho_{pin} = 0$ and (F) $\rho_{pin} = 0.10$.}
\label{fig:4}
\end{figure*}

\begin{figure*}[!htpb]
  \centering
   \includegraphics[width=0.45\textwidth]{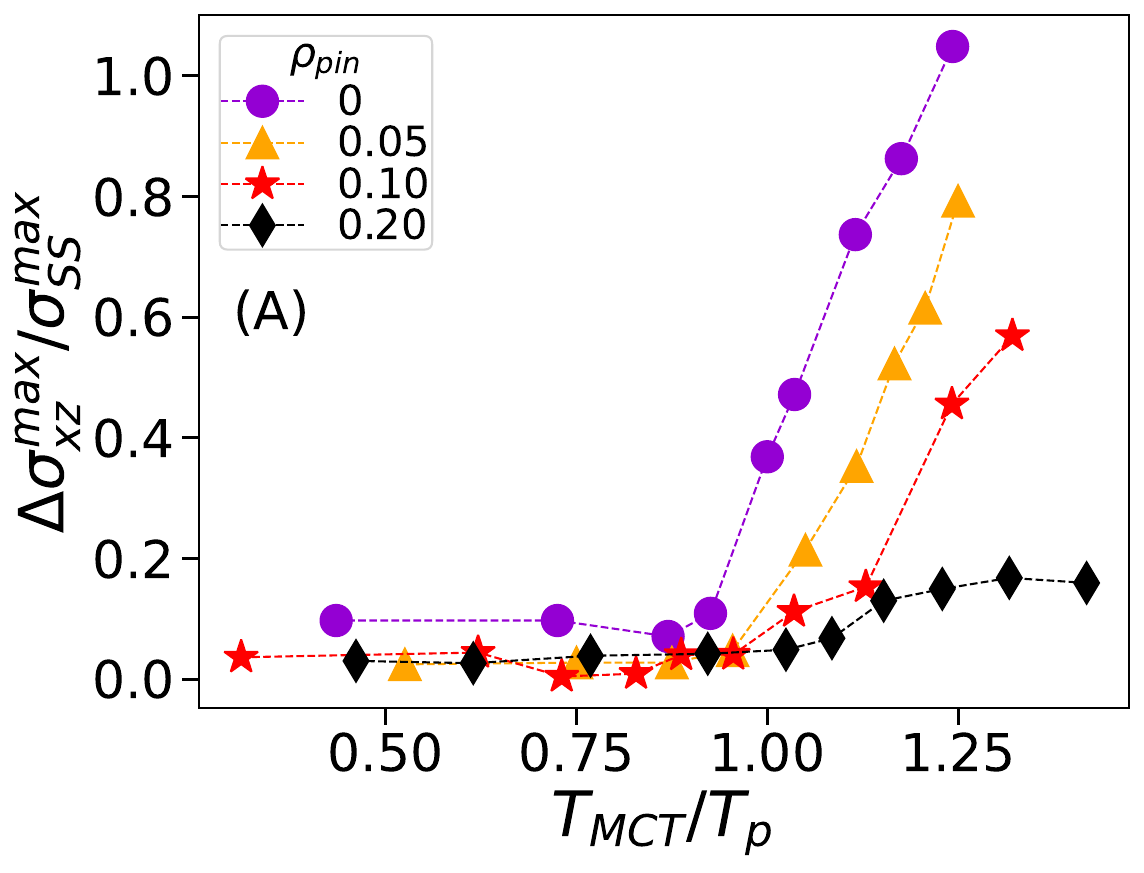}
   \includegraphics[width=0.45\textwidth]{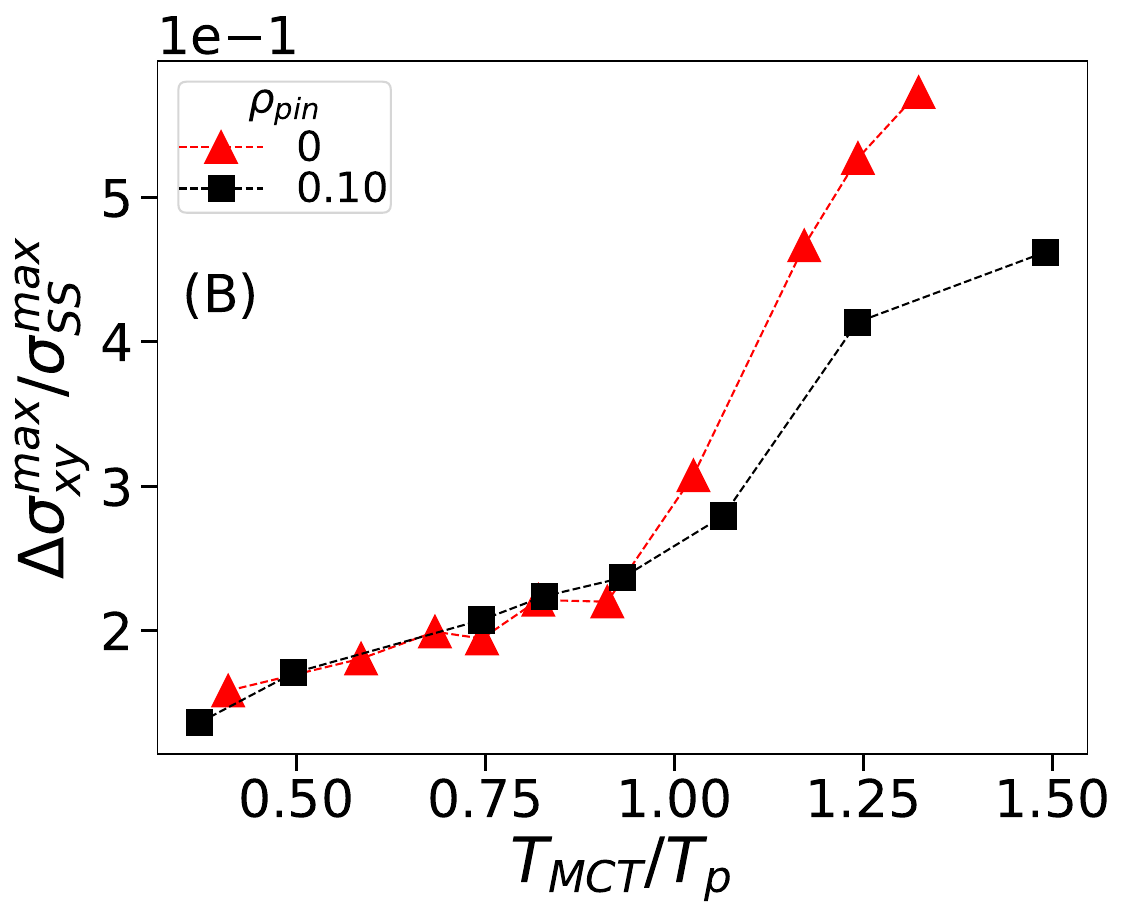}
  \caption{(A) Maximum steady state stress drop $\Delta\sigma^{max}_{xz}$ at yielding is plotted against parent temperature for different pin concentration $\rho_{pin}$ in 3D. Stress drop is normalized by steady-state maximum stress $\sigma^{max}_{SS}$, and temperature is scaled by $T_{MCT}$ for better comparison. It is obvious that at the same degree of annealing with increasing pin concentration, the stress drop decreases, indicating a more ductile-like feature for the pinned system. A similar plot for 2D is shown in (B).}
\label{fig:7}
\end{figure*}

\begin{figure*}[htpb]
  \centering
\includegraphics[width=0.9\textwidth,height=0.7\textwidth]{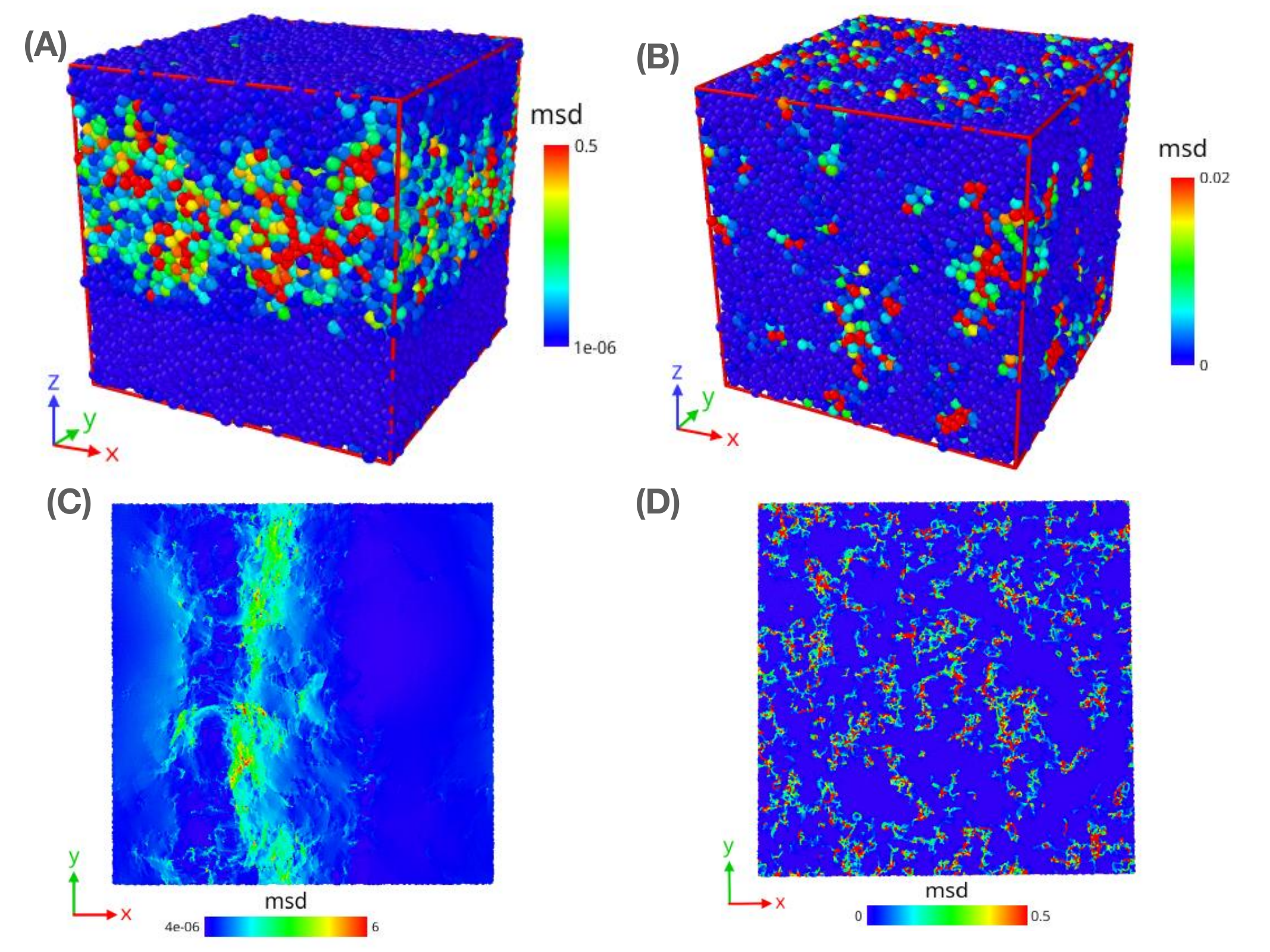}
  \caption{Shear band is shown for pin concentration (A) $\rho_{pin} = 0$ ($T_p = 0.42$ sheared at $\gamma_{max} = 0.09$), (B) $\rho_{pin} = 0.20$ ($T_p = 0.80$ sheared at $\gamma_{max} = 0.165$) in 3D. Shear band in 2D is also shown in (C) $\rho_{pin} = 0$ ($T_p = 0.35$ sheared at $\gamma_{max} = 0.07$) and (D) $\rho_{pin} = 0.10$ ($T_p = 0.50$ sheared at $\gamma_{max} = 0.15$). For all pin concentrations, $T_p$ is chosen well below their respective $T_{MCT}$. Samples are sheared at their yielding amplitude $\gamma_y$. It is clear that the shear band disappears in the pinned system in 3D as well as in 2D.}
\label{fig:5}
\end{figure*}

\begin{figure*}[htpb]
   \includegraphics[width=0.325\textwidth]{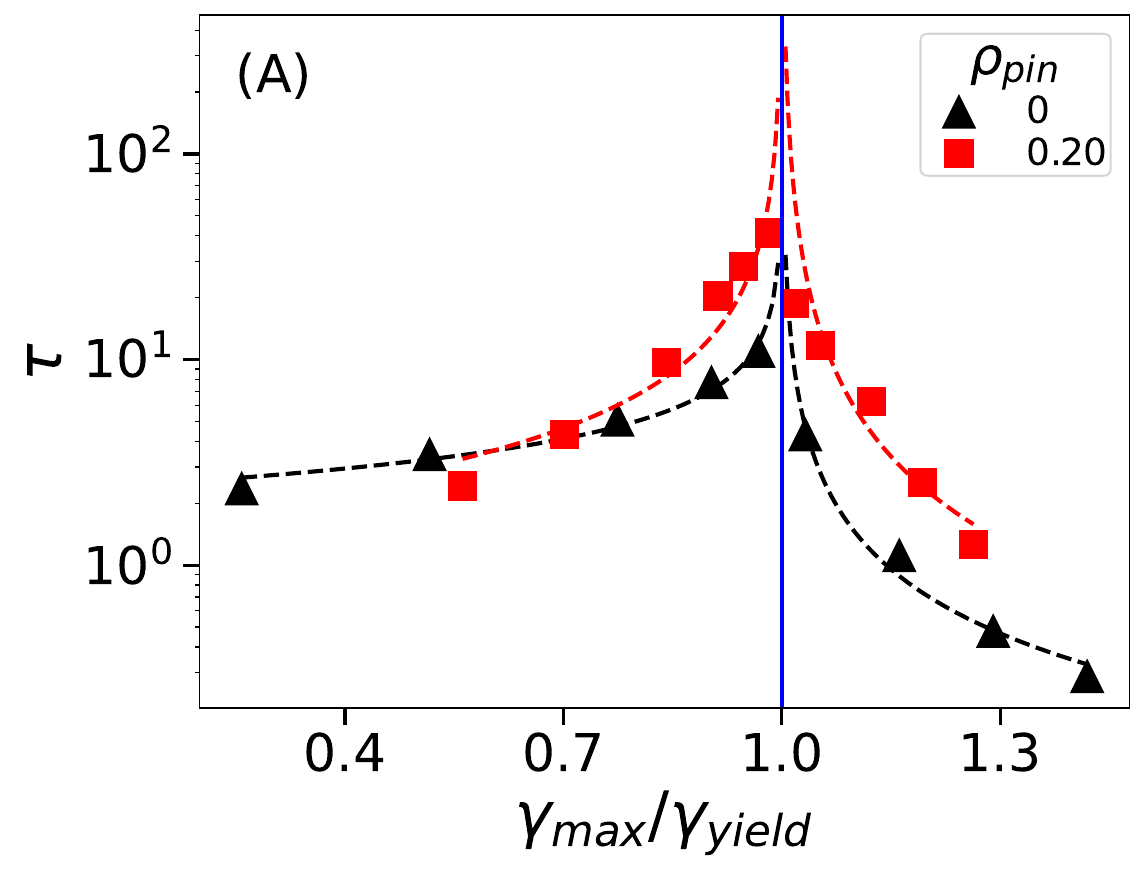}
   \includegraphics[width=0.325\textwidth]{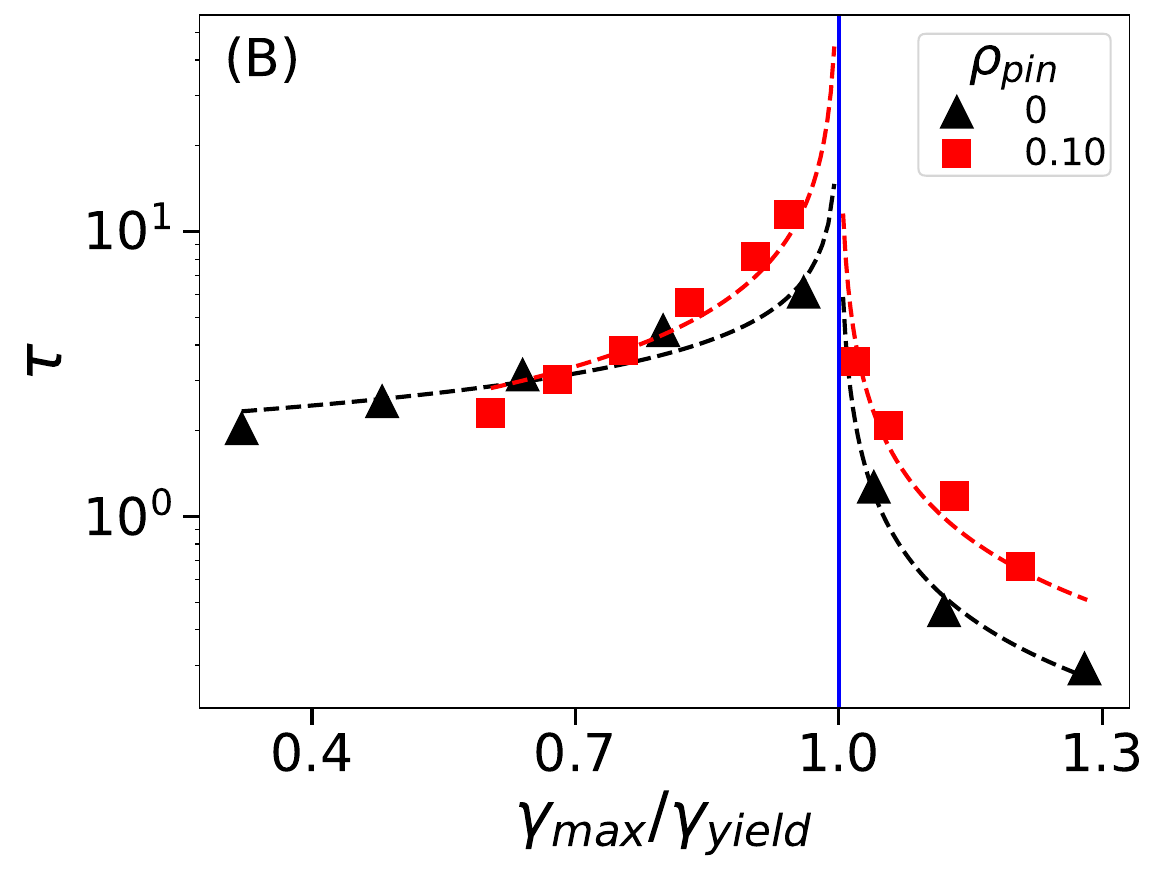}
   \includegraphics[width=0.325\textwidth]{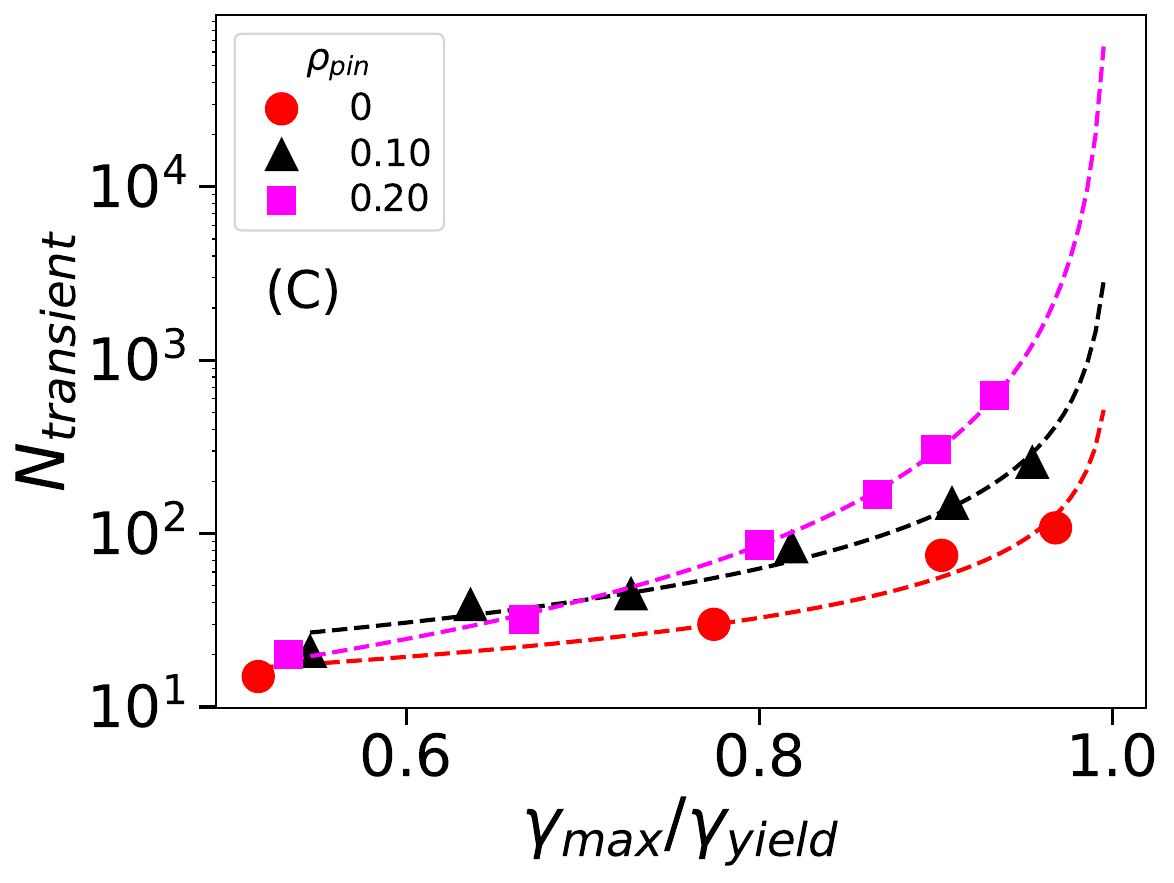}
   \includegraphics[width=0.325\textwidth]{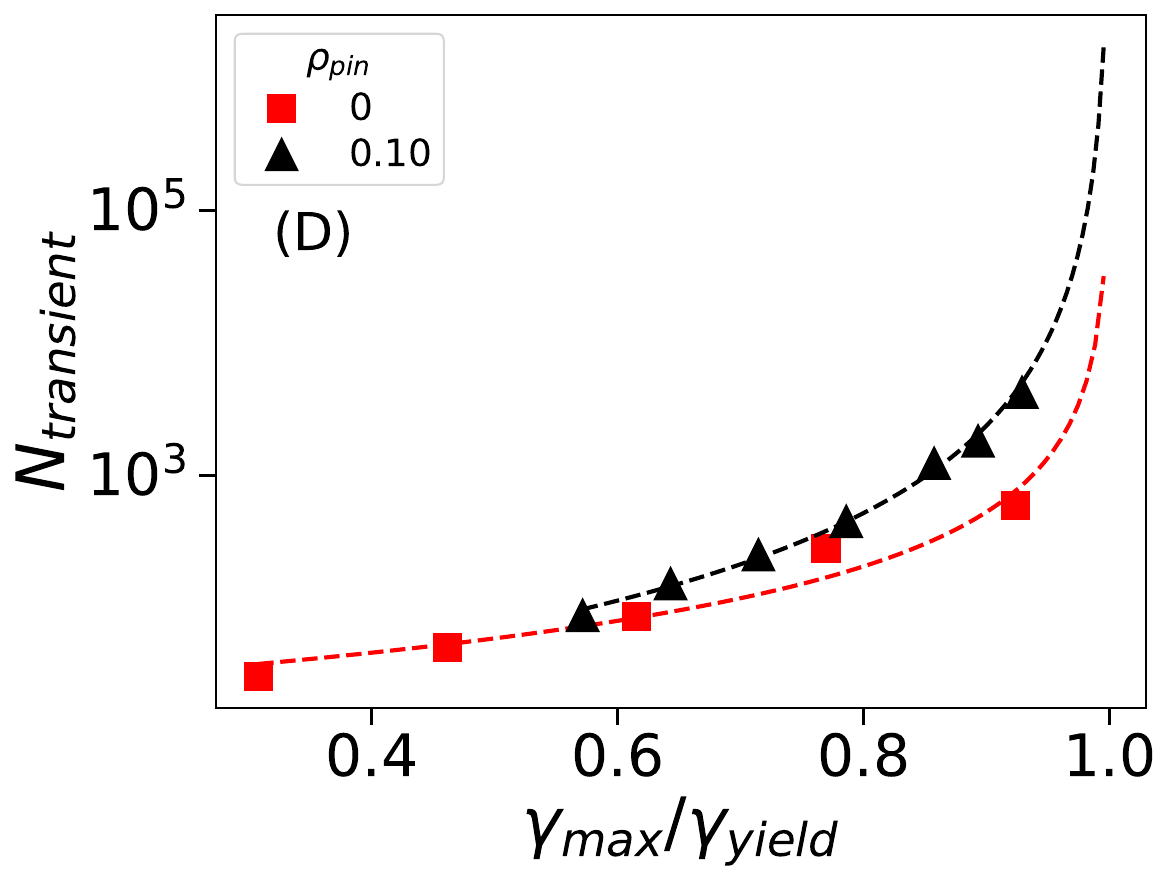}
   \includegraphics[width=0.325\textwidth]{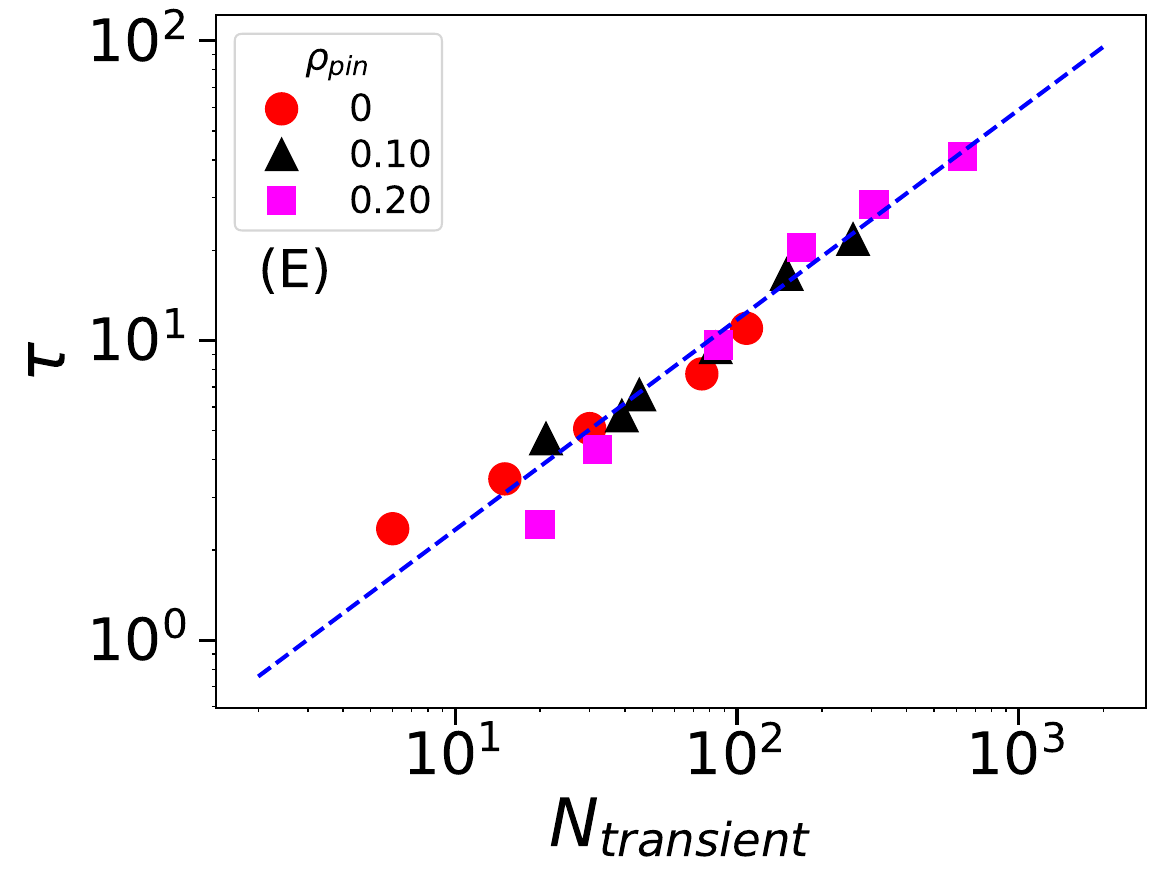}
   \includegraphics[width=0.325\textwidth]{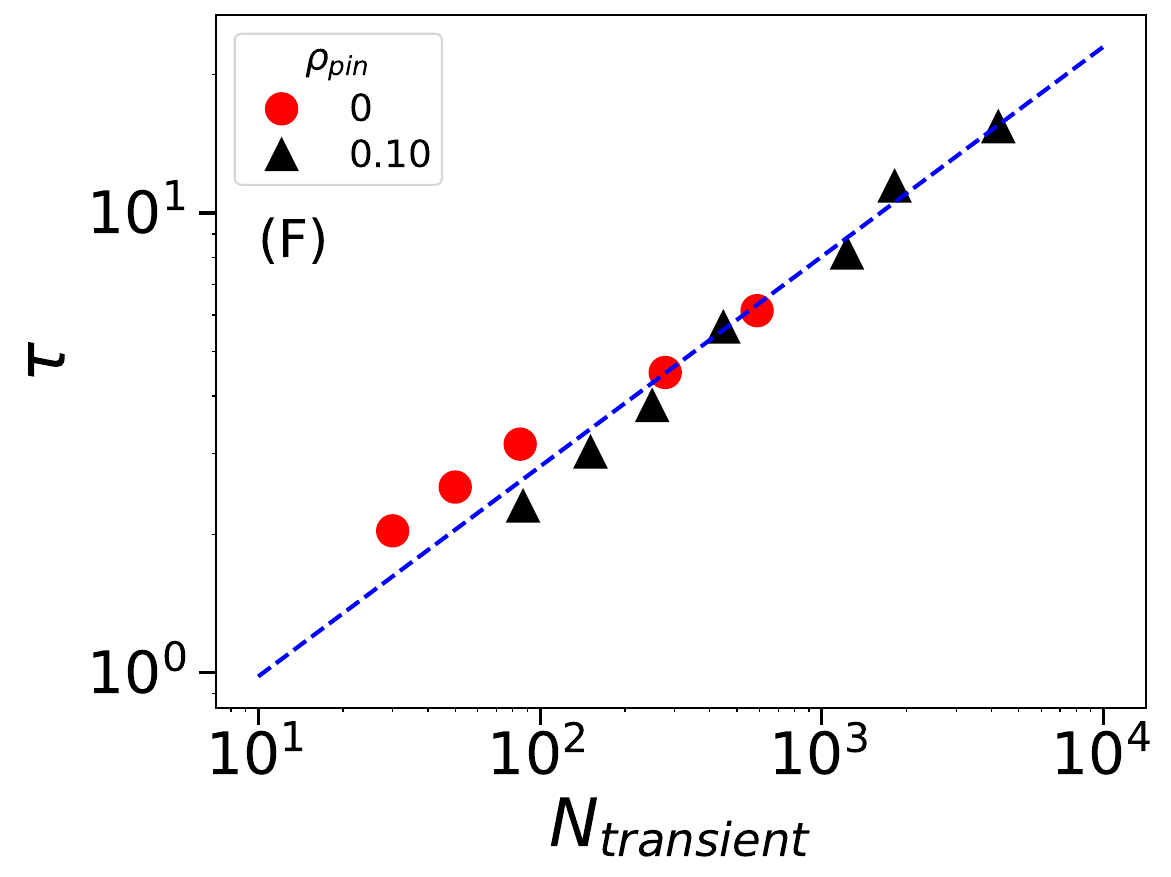}
  \caption{Timescale $\tau$ to reach the steady state is plotted against $\gamma_{max}$ normalised by $\gamma_{yield}$ for different pin concentrations in (A) 3D and (B) 2D. Data points are fitted via power law : $\tau = a\left|1 - \frac{\gamma_{max}}{\gamma_{yield}}\right|^{-b}$. The solid vertical blue line corresponds to $\gamma_{max} = \gamma_{yield}$. It is clearly seen that the timescale diverges at $\gamma_{max} \approx \gamma_y$. Moreover, the rate of increase of $\tau$ with $\gamma_{max}/\gamma_{yield}$ is much higher near yielding for the pinned system than $\rho_{pin} = 0$ system in 3D as well as in 2D. The Number of plastic drops $N_{transient}$ throughout the transient cycles (before reaching steady state) is plotted for different pin concentrations $\rho_{pin}$ in (C) 3D and (D) 2D. This indicates that pinned systems go through a significantly larger number of plastic drops than unpinned systems in both dimensions. Timescale $\tau$ vs plastic drops $N_{transient}$ is shown in (E) 3D and (F) 2D. Data points are fitted via power law: $\tau \propto N^\beta_{transient}$, where $\beta \approx 0.70$ and 0.45 for 3D and 2D, respectively.}
\label{fig:6}
\end{figure*}

\SK{After equilibrating the samples at the parent temperature, energy minimization was performed to obtain the amorphous solid state. For each sample, a large number of shear cycles were then applied using Athermal Quasistatic (AQS) shear with varying amplitudes until steady states were achieved. Energy as a function of cycle number for different strain amplitudes and temperatures is shown in the Supplementary Information. In Fig.\ref{fig:2}, the steady-state energy $E_{SS}$ is plotted as a function of $\gamma_{max}$ for different parent temperatures $T_p$ and pinning fractions. Panels (A)-(D) show data for 3D with pinning fractions of $0.0$, $0.05$, $0.10$, and $0.20$ respectively, while panels (E) and (F) present data for 2D with pinning fractions of $0.0$ and $0.10$ respectively. In Fig.\ref{fig:2}(A), we present data for the unpinned system. Consistent with previous observations \cite{bhaumik2021role, chatterjee2024role}, we find that all poorly annealed glasses ($T_p > 0.435$, where $T = 0.435$ denotes $T_{\text{MCT}}$) exhibit mechanical annealing up to $\gamma_{\text{max}} = 0.075$; beyond this point, $E_{\text{SS}}$ begins to increase, indicating all these poorly annealed glass have a common yield point referred to $\gamma_y = \gamma_c \approx 0.075$. In contrast, well-annealed samples ($T_p \leq 0.435$) do not show mechanical annealing with increasing $\gamma_{\text{max}}$ before yielding. However, after yielding, these well-annealed samples display a discontinuous energy jump. The yield point in well-annealed samples is larger than in poorly annealed ones ($\gamma_y > \gamma_c$), and with an increasing degree of annealing, the yield point $\gamma_y$ increases progressively.}  

\SK{However, a different scenario emerges for the pinned liquid. With increasing pinning fraction, the yielding amplitude for poorly annealed samples rises, and, as in the unpinned system, all poorly annealed glasses yield at nearly the same point. However, as annealing increases, the yield point shifts to larger values, though this shift is smaller than in the unpinned system. Moreover, as the pinning fraction increases, the shift in yield point becomes progressively smaller. This observation is consistent with recent findings that the fragility of the initial glass former influences the yielding behavior \cite{chatterjee2024role}. One key difference from previous observations is the change in steady-state energy within the diffusive region above the yielding point. As the pinning fraction increases, this difference in steady-state energy also grows, indicating a trend that warrants further investigation to understand the underlying mechanisms.  Next, we examine whether this behavior is consistent across dimensions by studying two different liquids with pinning fractions of $0\%$ and $10\%$ in 2D. We observe similar behavior as in 3D: with increasing pinning fraction and consequently decreasing fragility, the yielding behavior changes in a similar manner to the 3D case. This consistency suggests that the observed behavior is indeed universal.}

\SK{In Fig.\ref{fig:3}, we plot the yielding amplitude, $\gamma_{\text{yield}}$, as a function of temperature for different pinning fractions. The inset of panel (A)  clearly demonstrates that the yielding amplitude remains unchanged with sample age when samples are poorly annealed ($T_p > T_{\text{MCT}}$). However, it begins to increase as the degree of annealing improves below $T_{\text{MCT}}$. We denote the yielding amplitude of poorly annealed samples as $\gamma_{c}$. In the main panel of Fig.\ref{fig:3}, we show $\gamma_{\text{yield}}$ scaled by $\gamma_{c}$ as a function of temperature scaled by $T_{\text{MCT}}$. We observe that, although poorly annealed glass exhibits similar yielding behavior across different pinning fractions, the behavior changes markedly in the well-annealed regime. With increasing pinning fraction, the relative change in yielding amplitude becomes smaller, further confirming the universal nature of the relationship between yielding behavior and fragility \cite{chatterjee2024role}.}


\SK{Next, to understand this yielding behavior, we have plotted the steady state maximum stress $\sigma^{max}_{xz}$ at different $\gamma_{max}$ for a wide range of temperature $T_p$ for unpinned ($\rho_{pin} = 0$) system in 3D in Fig.\ref{fig:4}(A). $\sigma^{max}_{xz}$ increases with $\gamma_{max}$ till yield point, and after yielding, it starts to decrease. The well-annealed samples show a sharp drop of stress at yielding, indicating brittle-like features. For the pinned systems in Fig.\ref{fig:4}(B) - (D), as the pin concentration $\rho_{pin}$ increases, we see the stress drop decreases, indicating more ductile-like behavior. This suggests that brittle to ductile features can be controlled by fragility, as also seen in recent work \cite{chatterjee2024role}. Moreover, the difference in steady-state stress after yielding increases with pin concentration $\rho_{pin}$. Similar characteristics are also observed in 2D at $\rho_{pin} = 0$ and $0.10$ in Fig.\ref{fig:4}(E) and (F) respectively. The stress drops $\Delta\sigma^{max}_{xz}$ at the yield strain defined by the difference in stress between the peak stress and the steady-state stress normalized by the steady-state stress $\sigma^{max}_{SS}$ is plotted against the degree of annealing quantified parent temperature, $T_p$ for different pin concentration $\rho_{pin}$ in Fig.\ref{fig:7}(A) and (B) respectively for 3D and 2D. These clearly show that with increasing pinning concentration, the yielding is becoming more rounded.}

\vskip 0.05in
\noindent{\bf Structure of Shear Bands: }
\SK{To study the difference in shear band structure in pinned and unpinned systems, we have used system size $N = 50000$. We have prepared samples at $T_p = 0.42$ for $\rho_{pin} = 0$ and $T_p = 0.80$ for $\rho_{pin} = 0.20$ in 3D. Similarly we equilibrated samples in case of 2D at $T_p = 0.35$ for $\rho_{pin} = 0$ and $T_p = 0.50$ for $\rho_{pin} = 0.10$. For each system, the temperatures are chosen well below their respective $T_{MCT}$. We have simulated $5$ independent samples in each case. To get the shear band, we sheared the samples above their yielding amplitude till the steady state was reached and then took configurations at the end of two consecutive cycles in the steady state and calculated the mean-squared-displacement (MSD) of each particle. We highlighted the particles that move larger than a chosen cut-off distance during one single cycle in the steady state by color coding in OVITO software. From Fig.\ref{fig:5}, we see the shear band appears in $\rho_{pin} = 0$ systems but disappears in all the pinned systems in 3D as well as in 2D.}

\vskip 0.05in
\noindent{\bf Microscopic understanding of timescale:}
\SK{We further studied the time $\tau$ needed for a system to reach the steady state sheared at an amplitude $\gamma_{max}$. The timescales are extracted from the stretched exponential fitting of the stroboscopic energy vs. the number of shear cycles graphs shown in SI. We have plotted $\tau$ against $\gamma_{max}$ normalised over $\gamma_{yield}$ for better presentation in Fig.\ref{fig:6}(A) and (B) for 3D and 2D respectively. The blue solid line denotes $\gamma_{max} = \gamma_{yield}$ and we see the timescale $\tau$ diverges as $\gamma_{max} \approx \gamma_{yield}$. We fitted the data points with power law, $\tau = a\left|1 - \frac{\gamma_{max}}{\gamma_{yield}}\right|^{-b}$. We found the exponent $b$ (for both before and after yielding branches) is higher for pinned systems compared to unpinned systems in 3D as well as in 2D. We got $b = 0.48$ and $0.90$ for the before-yielding branch and for the after-yielding branch $b = 1.03$ and 1.35 in $\rho_{pin} = 0$ and 0.20 respectively in 3D. In case of 2D $b = 0.37$ and 0.63 for before yielding branch and for after yielding branch $b = 0.76$ and 1.01 in $\rho_{pin} = 0$ and 0.10 respectively. It indicates that near yielding, at a given $\gamma_{max}/\gamma_{yield}$, pinned system (strong glass-former) needs a significantly large number of cycles to reach the steady state compared to unpinned system (fragile glass-former). To understand the microscopic point of view behind this timescale analysis, we plotted the number of plastic drops ($N_{transient}$) throughout the transient cycles (before reaching the steady state) against $\gamma_{max}$ normalized by yield point $\gamma_{yield}$ in 3D and 2D in Fig.\ref{fig:6}(C) and (D) respectively. We see that the pinned system goes through a significantly larger number of plastic drops than the unpinned ($\rho_{pin} = 0$) system at a given $\gamma_{max}/\gamma_{yield}$. We plotted the correlation between timescale $\tau$ and number of plastic drops $N_{transient}$ in 3D and 2D in Fig.\ref{fig:6}(E) and (F), respectively. We fitted the data with power law $\tau \propto N^\beta_{transient}$, where $\beta = 0.70$ and 0.45 in the case of 3D and 2D, respectively. A universal relation between the time required to reach the steady state and the number of plastic drops the system goes through suggests a deeper connection between the underlying energy landscape and the searching process the system goes through during the oscillatory shear cycles. A possible microscopic understanding using the framework proposed in \cite{mungan2019structure,mungan2019networks,mungan2019cyclic,terzi2020state,adhikari2023encoding} of a network of energy minima and how the system traverses through the network might give us the origin of this universal power-law relation. This might also give us clues about how one can use the oscillatory shear response of the system to understand the network topology of the energy landscape. A better understanding of why strong glass-formers explore through a significantly larger number of plastic drops compared to their fragile counterparts before they find the absorbing states under oscillatory shear will certainly be very useful to even understanding the origin of fragility and its variation in various glass-formers.}

\vskip 0.05in
\noindent{\bf Conclusion: }
\SK{In this study, we investigated the yielding behavior of randomly pinned systems with varying degrees of annealing under oscillatory shear deformation. Our primary objective was to determine whether recent observations of the effect of changing fragility on yielding behavior are universal or not. Previous studies on sphere assemblies with harmonic interactions revealed that changing density, and thus fragility, can impact the nature of yielding behavior. It was found that poorly annealed glasses exhibit similar behavior across different fragile glass formers. In contrast, well-annealed glasses display a significant change in the nature of yielding. Specifically, the yielding amplitude for strong glasses remains nearly the same regardless of annealing, whereas, for fragile glasses, the yielding amplitude increases significantly and becomes more brittle as annealing increases.}

\SK{Although, by varying the fraction of pinned particles, one can tune the fragility of the system, this model differs significantly from previous ones. Unlike earlier models with only repulsive interactions, this system includes attractive and repulsive forces. Furthermore, previous studies on randomly pinned systems under uniform shear revealed distinct yielding mechanisms between pinned and unpinned systems. Specifically, unpinned systems exhibit shear band formation, whereas the pinned systems display homogeneous yielding without shear bands. Therefore, it is not immediately obvious that the behavior observed in soft sphere harmonic systems will directly apply to our pinned system. To explore universality further, we investigated the model in both 2D and 3D dimensions.}

\SK{We observed that the unpinned system, which is a fragile glass former, exhibits the expected behavior: all poorly annealed glasses yield at the same point, but with increasing annealing, yielding becomes catastrophic, and the yielding amplitude changes significantly. However, as the pinning concentration increases, making it a strong glass former, the changes in yielding amplitude become smaller, and the yielding becomes more gradual. A similar behavior is also observed in 2D systems. These findings suggest that the observation is consistent with the soft sphere system, indicating the impact of fragility on yielding behavior is universal.}

\SK{Finally, in our investigation on the shear band formation under oscillatory shear deformation in the pinned system, we show that unlike unpinned systems, which exhibit clear vertical and horizontal shear bands, the pinned system does not display any shear band. Our study sheds light on the microscopic nature of yielding, suggesting that fragility plays a crucial role in determining the yielding mechanism regardless of the model or spatial dimensions. It will be interesting in the future to develop an elastoplastic model \cite{tyukodi2016finite} in which the effect of random pinning can be faithfully modeled, as such models will be very important to understanding the mechanical response of micro-alloyed materials at a scale that is amenable for practical applications.}

\vskip +0.05in
\begin{acknowledgments}
We acknowledge funding by intramural funds at TIFR Hyderabad from the Department of Atomic Energy (DAE) under Project Identification No. RTI 4007. SK acknowledges Swarna Jayanti Fellowship grants DST/SJF/PSA01/2018-19 and SB/SFJ/2019-20/05 from the Science and Engineering Research Board (SERB) and Department of Science and Technology (DST). Most computations are done using the HPC clusters procured using Swarna Jayanti Fellowship grants DST/SJF/PSA01/2018-19, SB/SFJ/2019-20/05. SK wants to acknowledge the research support from MATRICES Grant MTR/2023/000079 from SERB.
\end{acknowledgments}

\bibliography{apssamp}

\end{document}




\title[]{\Large \bf Effect of Random Pinning on the Yielding Transition of Amorphous Solid under Oscillatory Shear - Supplemental Information}

\author{Roni Chatterjee}
\affiliation{Tata Institute of Fundamental Research, 36/P, Gopanpally Village, Serilingampally Mandal, Ranga Reddy District, Hyderabad 500046, Telangana, India}
\author{Monoj Adhikari }
\affiliation{Tata Institute of Fundamental Research, 36/P, Gopanpally Village, Serilingampally Mandal, Ranga Reddy District, Hyderabad 500046, Telangana, India}
\author{Smarajit Karmakar}
\affiliation{Tata Institute of Fundamental Research, 36/P, Gopanpally Village, Serilingampally Mandal, Ranga Reddy District, Hyderabad 500046, Telangana, India}

\maketitle

\section{Stroboscopic Energy vs cycles}
\begin{figure}[H]
  \centering
   \includegraphics[width=0.49\textwidth,height=0.40\textwidth]{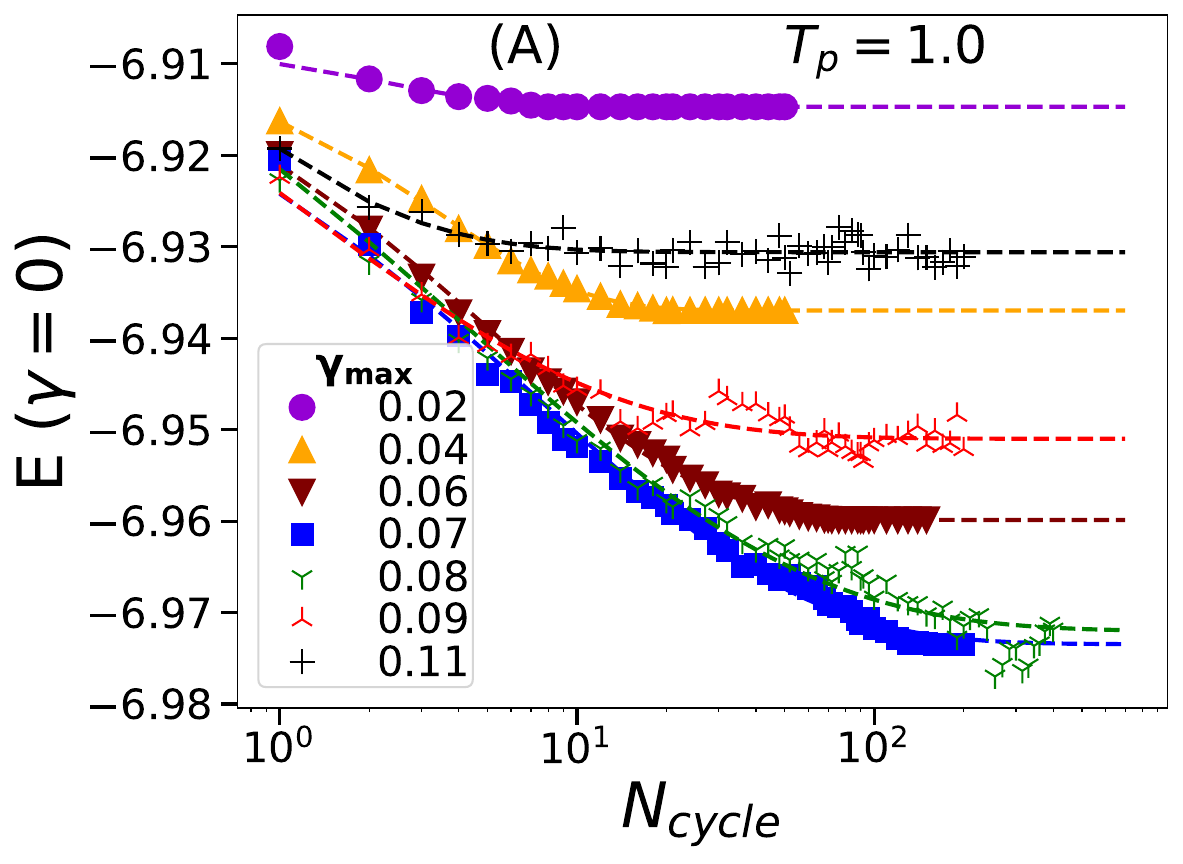}
   \includegraphics[width=0.49\textwidth,height=0.40\textwidth]{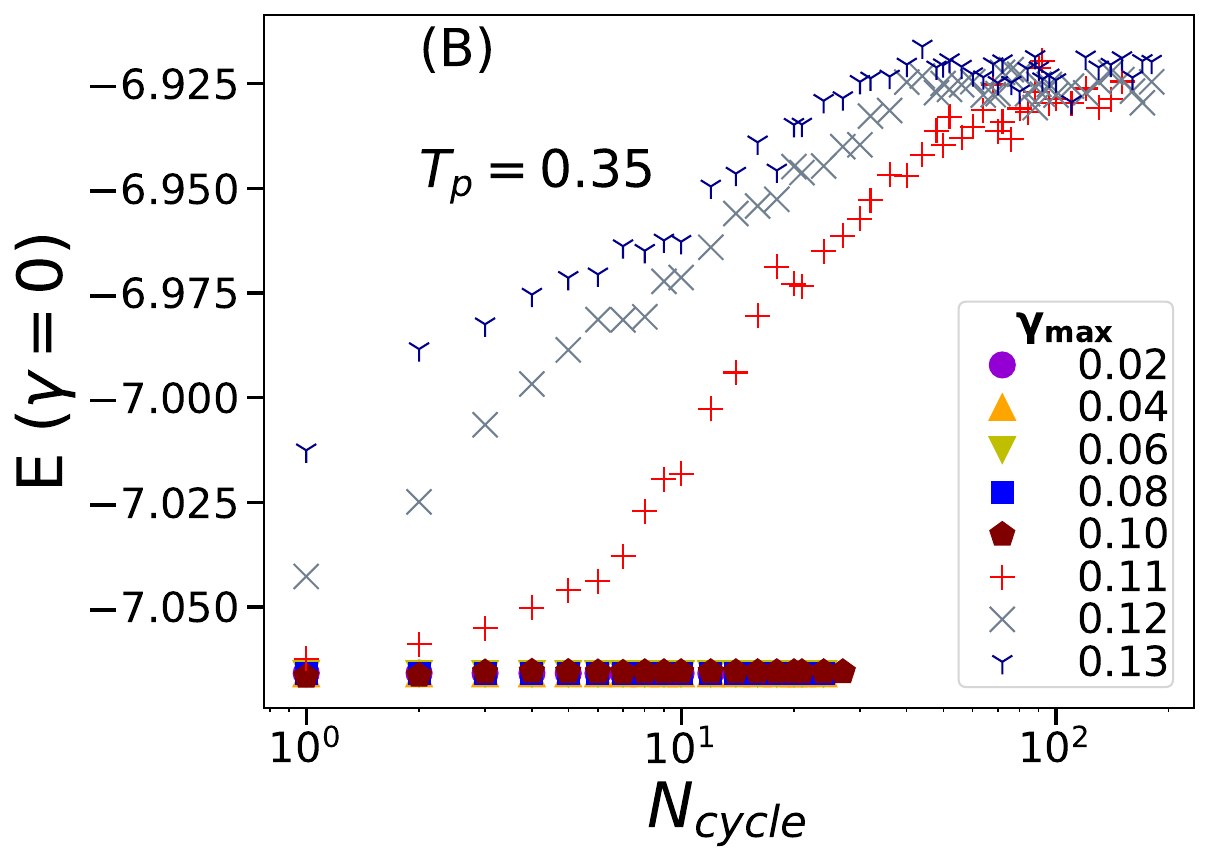}
  \caption{Plot of stroboscopic energy $E(\gamma = 0)$ vs number of shear cycles $N_{cycle}$ at different $\gamma_{max}$ for (A) poorly annealed ($T_p = 1.0$) and (B) well annealed ($T_p = 0.35$) samples at $\rho_{pin} = 0$ in 3D. Dashed lines represents the stretch exponential fitting to the data points. For poorly annealed sample yield point is at $\gamma_y \approx 0.075$ and well annealed sample has yield point at $\gamma_y \approx 0.105$.}
\label{fig:7}
\end{figure}

\begin{figure}[H]
  \centering
   \includegraphics[width=0.49\textwidth,height=0.40\textwidth]{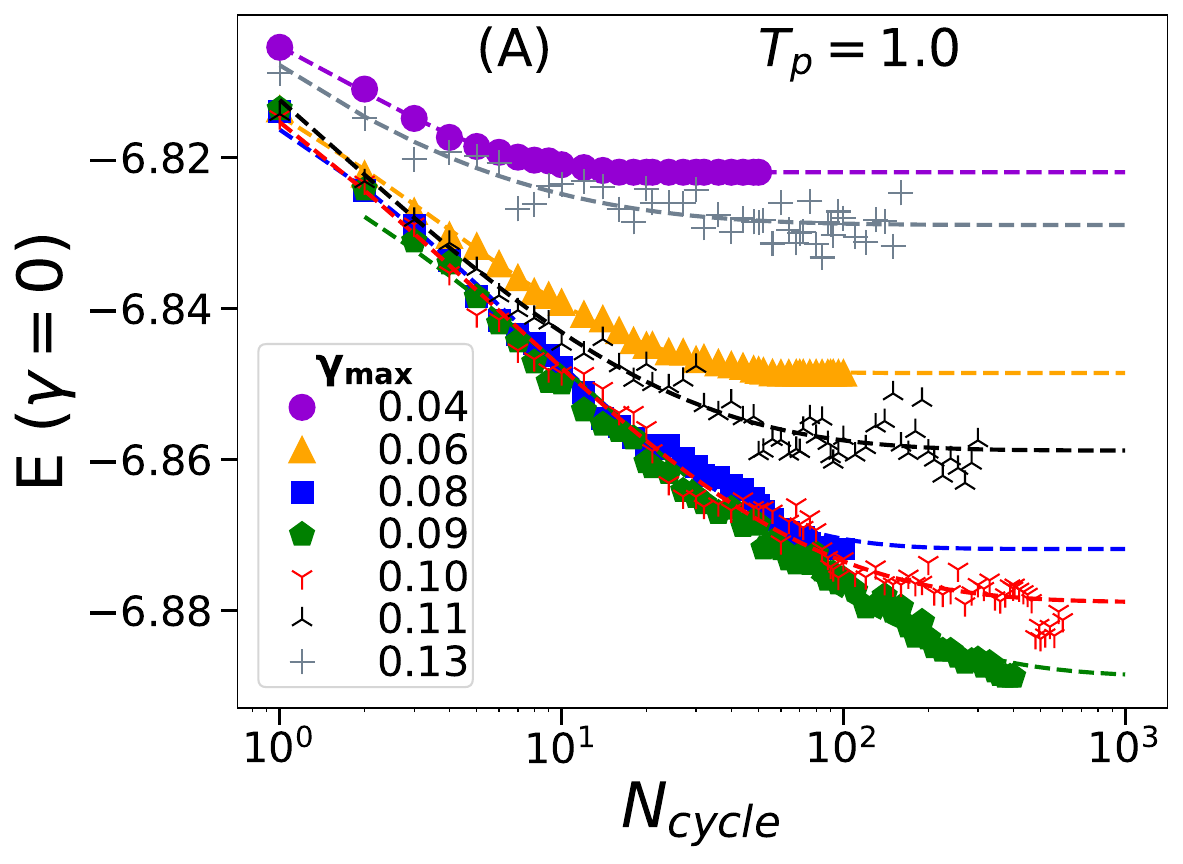}
   \includegraphics[width=0.49\textwidth,height=0.40\textwidth]{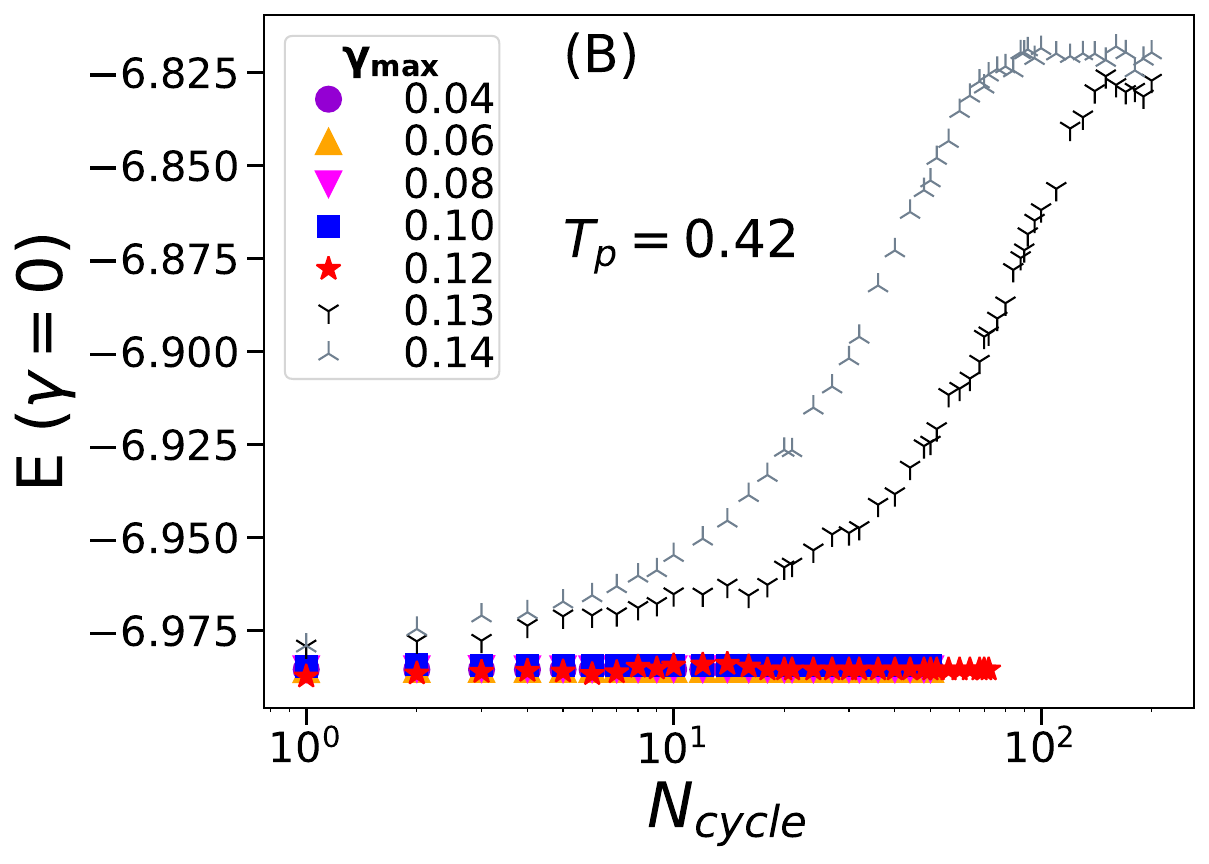}
 \caption{Plot of stroboscopic energy vs number of cycles $N_{cycle}$ at different $\gamma_{max}$ for (A) poorly annealed ($T_p = 1.0$) and (B) well annealed ($T_p = 0.42$) samples at $\rho_{pin} = 0.05$ in 3D. For poorly annealed sample yield point is at $\gamma_y \approx 0.095$ and well annealed sample has yield point at $\gamma_y \approx 0.125$.}
\label{fig:8}
\end{figure}

\begin{figure}[H]
  \centering
   \includegraphics[width=0.49\textwidth,height=0.40\textwidth]{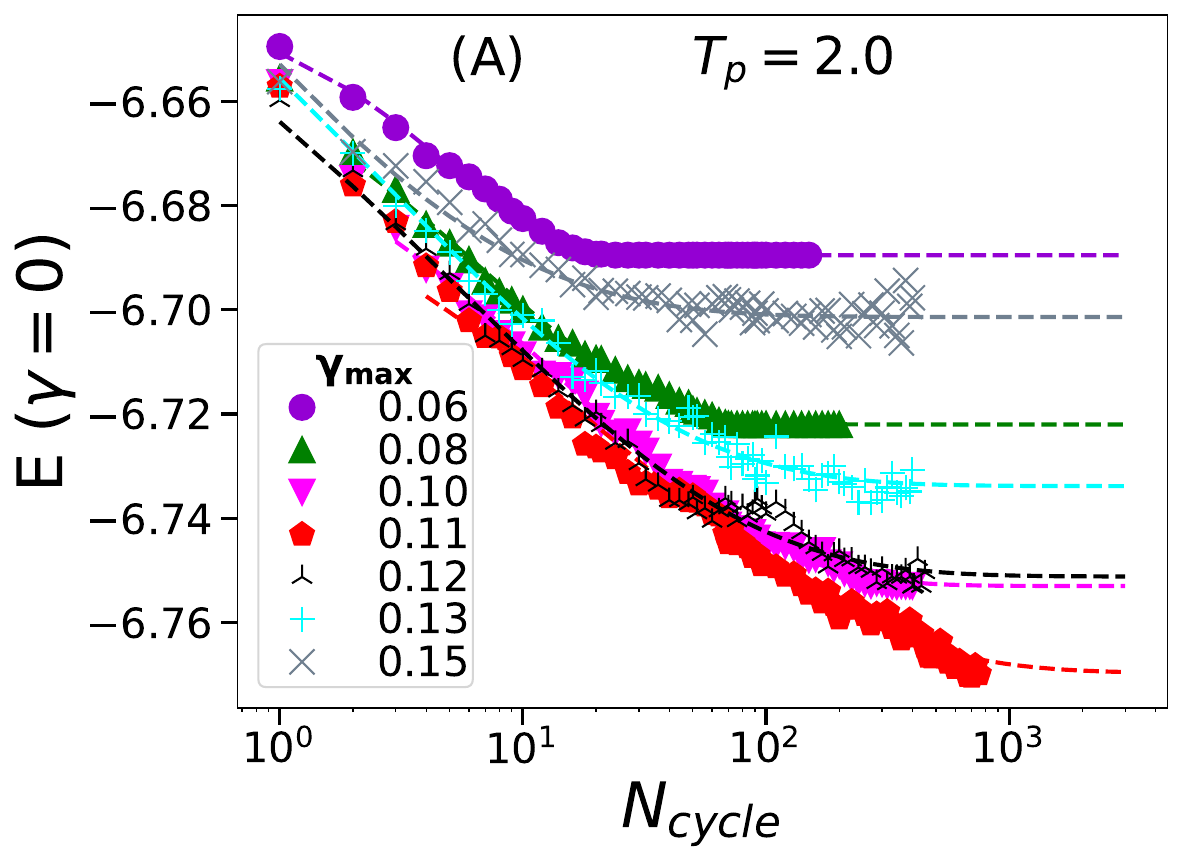}
   \includegraphics[width=0.49\textwidth,height=0.40\textwidth]{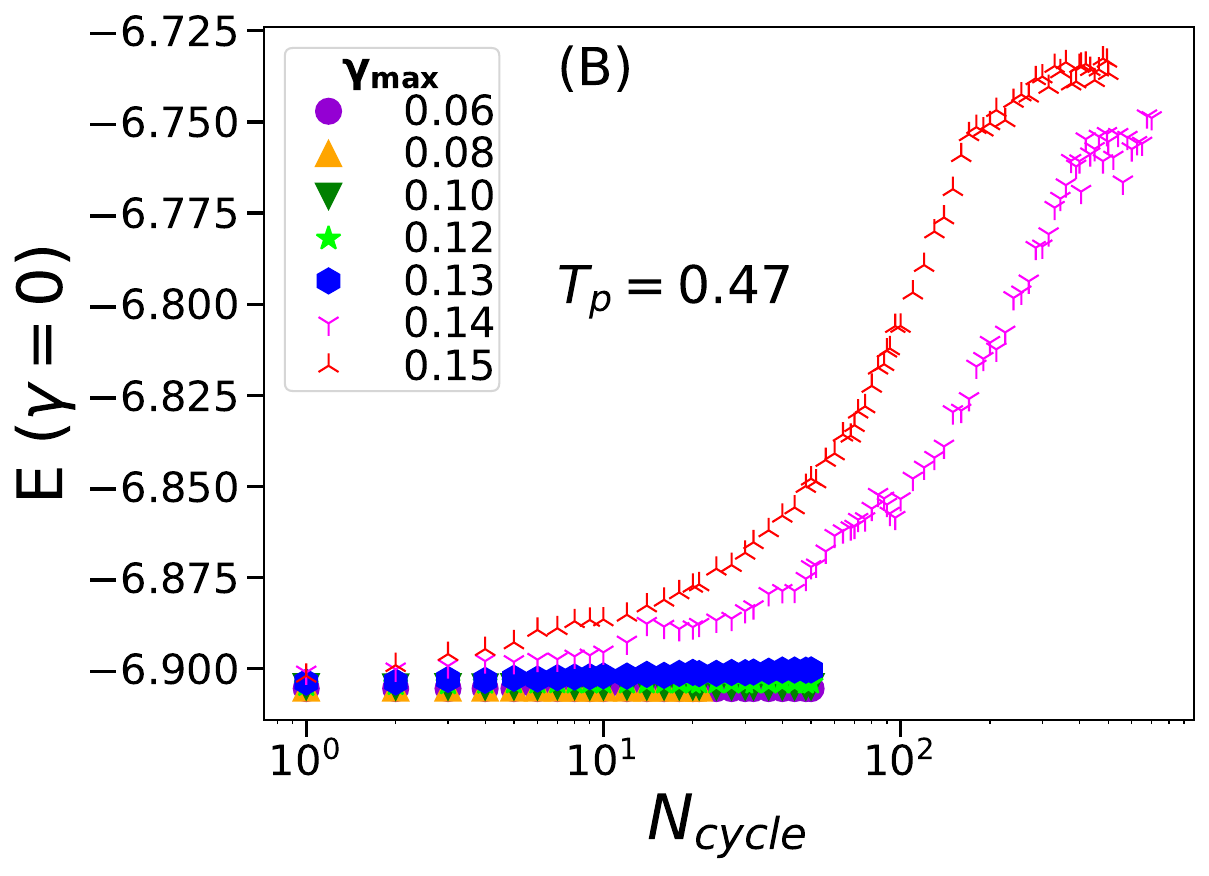}
 \caption{Plot of stroboscopic energy vs number of cycles $N_{cycle}$ at different $\gamma_{max}$ for (A) poorly annealed ($T_p = 2.0$) and (B) well annealed ($T_p = 0.47$) samples at $\rho_{pin} = 0.10$ in 3D. For poorly annealed sample yield point is at $\gamma_y \approx 0.11$ and well annealed sample has yield point at $\gamma_y \approx 0.14$.}
\label{fig:9}
\end{figure}

\begin{figure}[H]
  \centering
   \includegraphics[width=0.49\textwidth,height=0.40\textwidth]{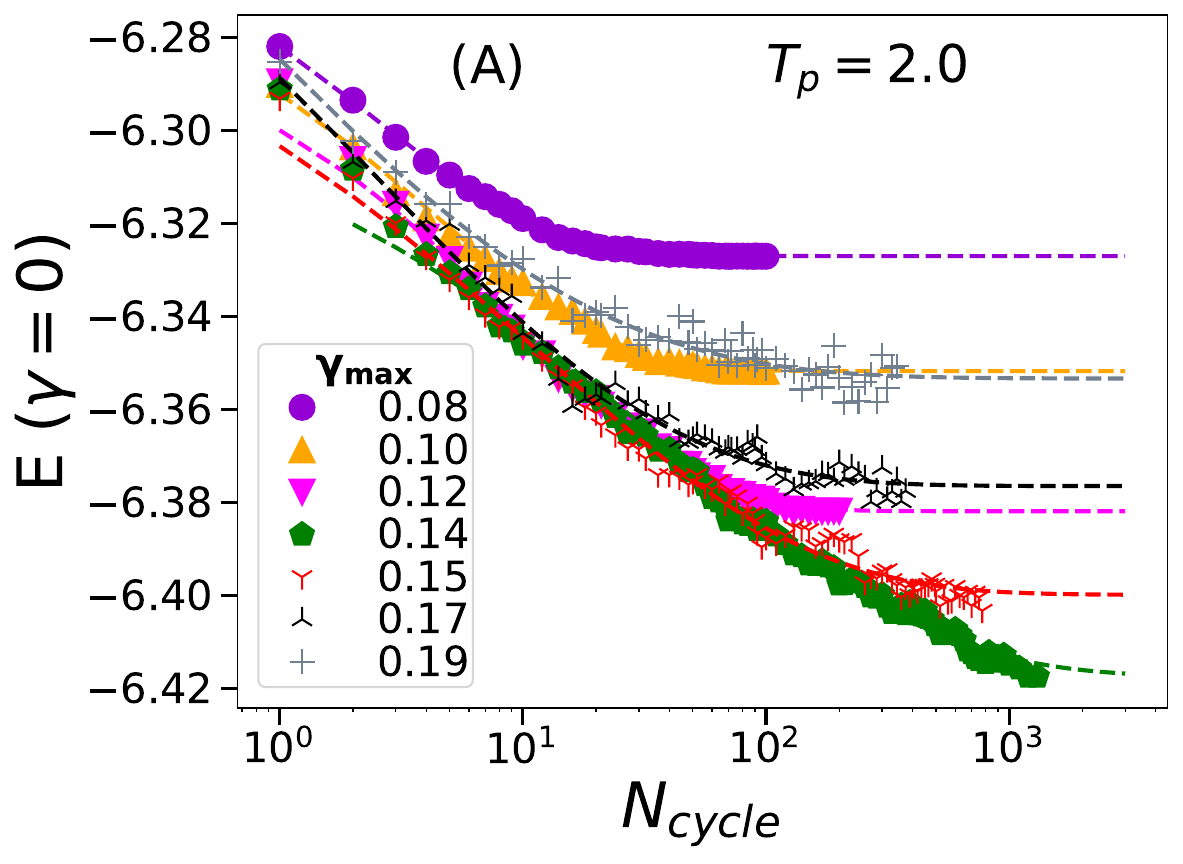}
   \includegraphics[width=0.49\textwidth,height=0.40\textwidth]{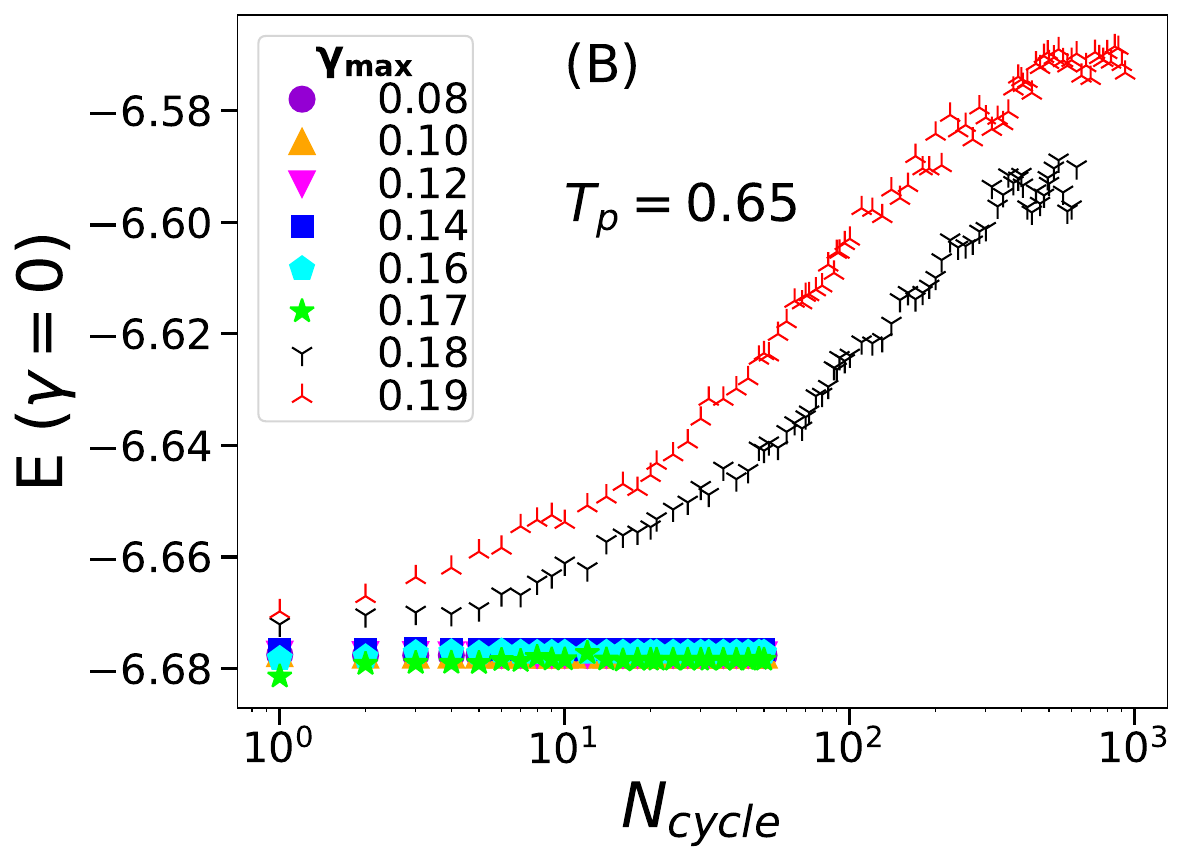}
 \caption{Plot of stroboscopic energy vs number of cycles $N_{cycle}$ at different $\gamma_{max}$ for (A) poorly annealed ($T_p = 2.0$) and (B) well annealed ($T_p = 0.65$) samples at $\rho_{pin} = 0.20$ in 3D. For poorly annealed sample yield point is at $\gamma_y \approx 0.15$ and well annealed sample has yield point at $\gamma_y \approx 0.18$.}
\label{fig:10}
\end{figure}

\begin{figure}[H]
  \centering
   \includegraphics[width=0.49\textwidth,height=0.40\textwidth]{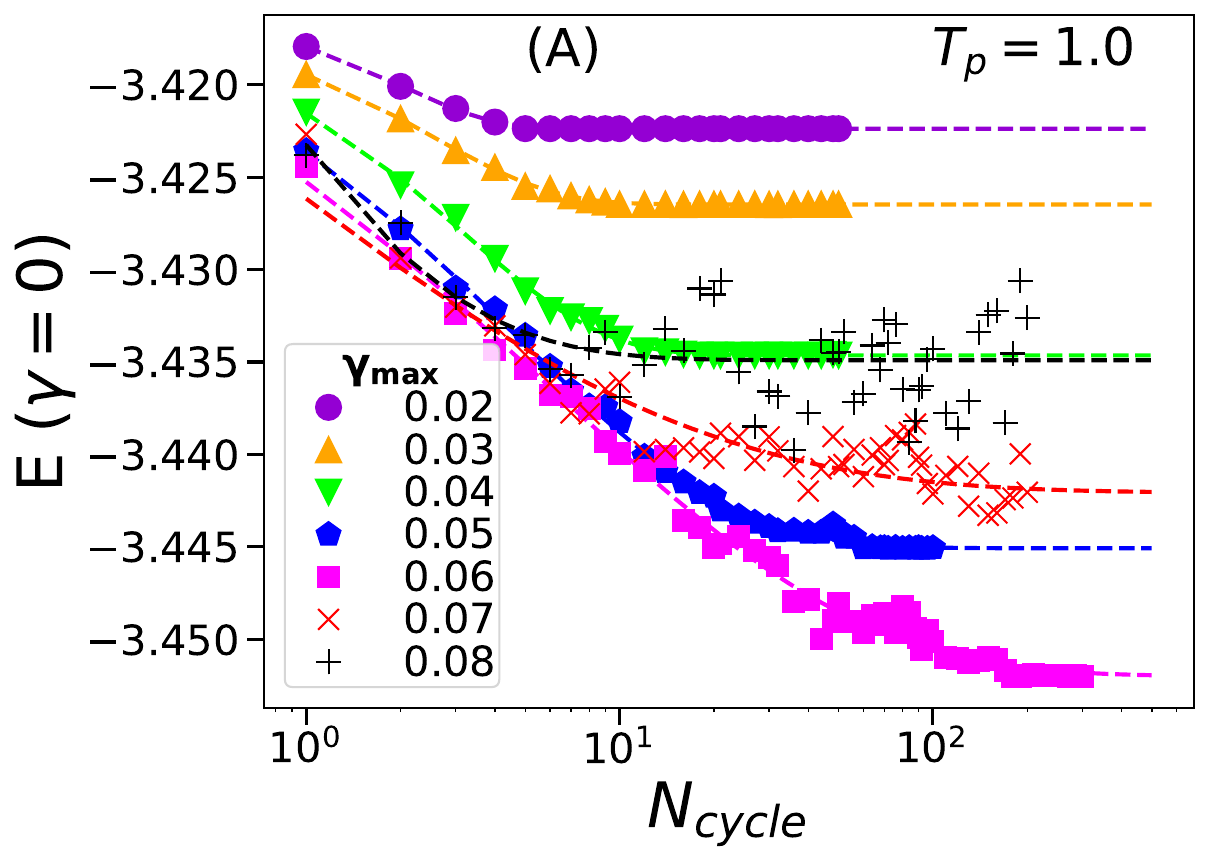}
   \includegraphics[width=0.49\textwidth,height=0.40\textwidth]{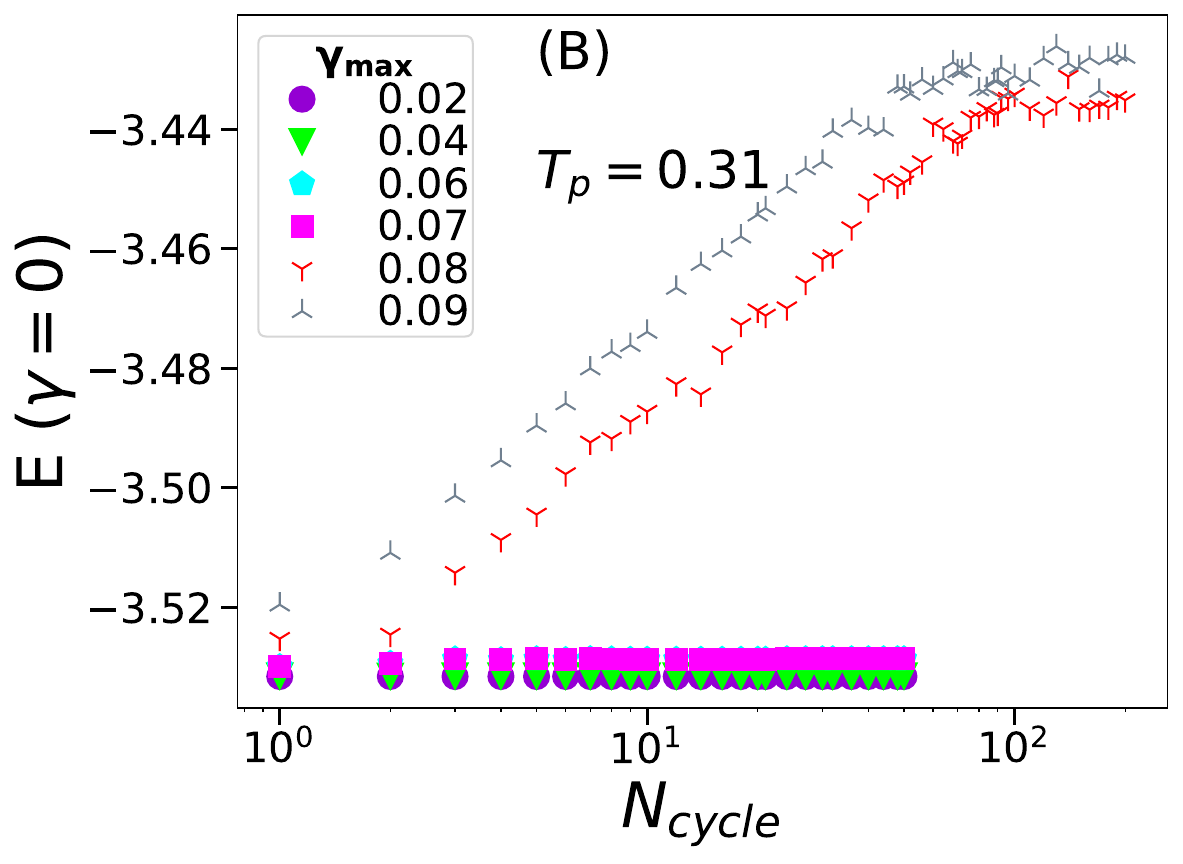}
 \caption{Plot of stroboscopic energy vs number of cycles $N_{cycle}$ at different $\gamma_{max}$ for (A) poorly annealed ($T_p = 1.0$) and (B) well annealed ($T_p = 0.31$) samples at $\rho_{pin} = 0$ in 2D. For poorly annealed sample yield point is at $\gamma_y \approx 0.065$ and well annealed sample has yield point at $\gamma_y \approx 0.072$.}
\label{fig:11}
\end{figure}

\begin{figure}[H]
  \centering
   \includegraphics[width=0.45\textwidth]{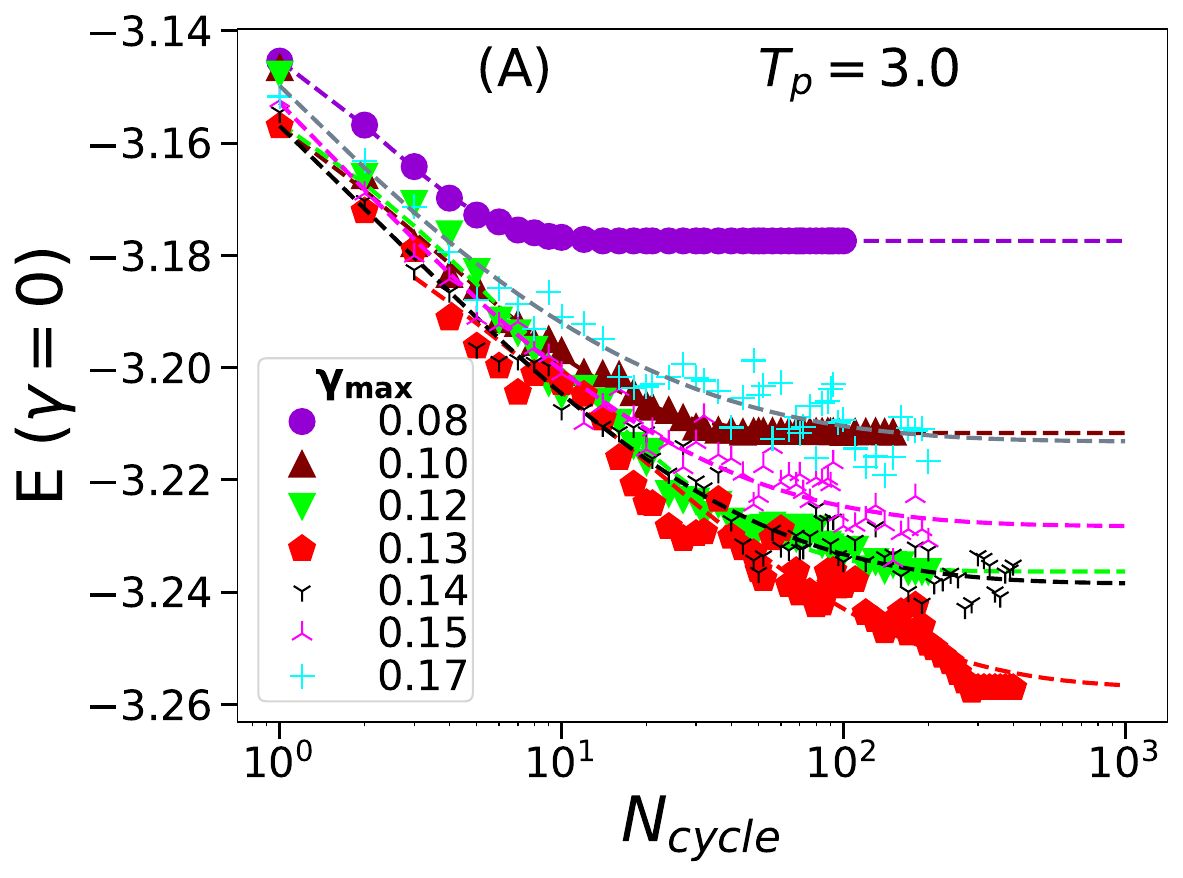}
   \includegraphics[width=0.45\textwidth]{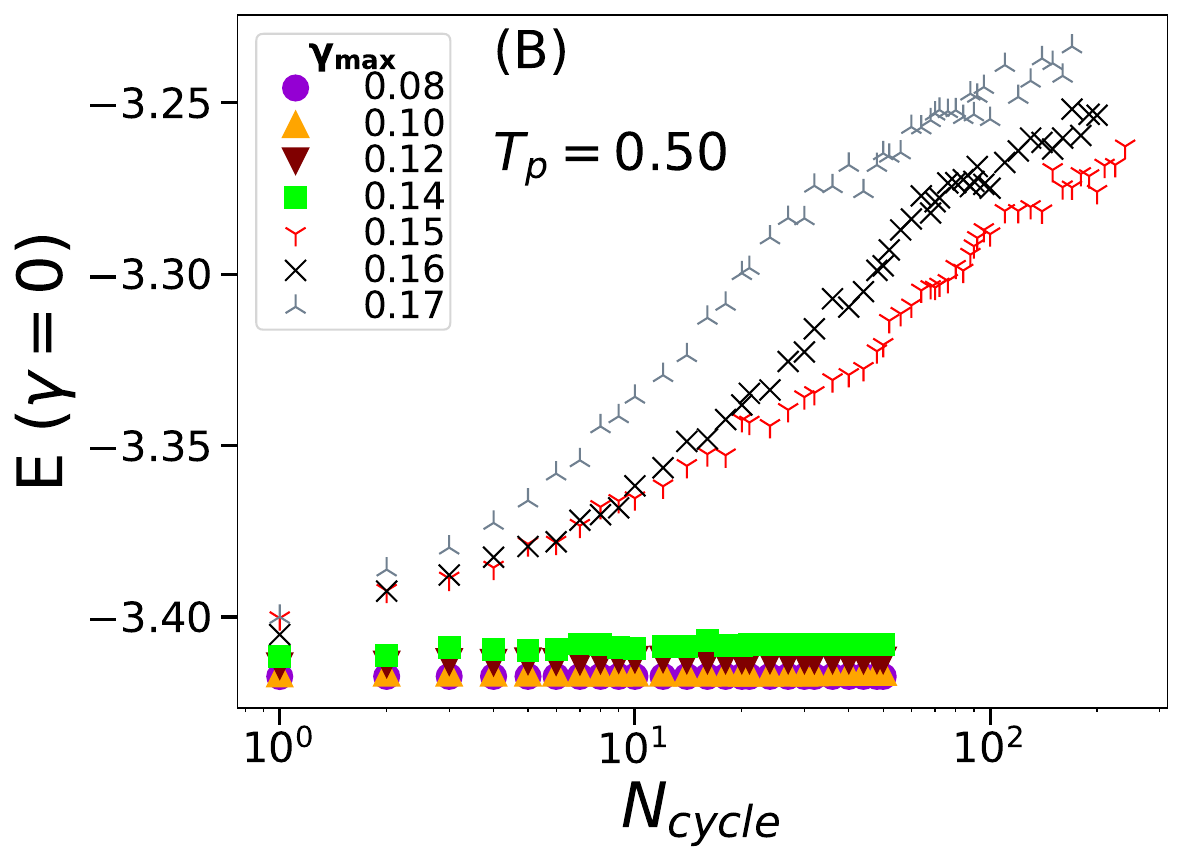}
 \caption{Plot of stroboscopic energy vs number of cycles $N_{cycle}$ at different $\gamma_{max}$ for (A) poorly annealed ($T_p = 3.0$) and (B) well annealed ($T_p = 0.50$) samples at $\rho_{pin} = 0.10$ in 2D. For poorly annealed sample yield point is at $\gamma_y \approx 0.135$ and well annealed sample has yield point at $\gamma_y \approx 0.146$.}
\label{fig:12}
\end{figure}

We have plotted the stroboscopic energy, $E(\gamma = 0$), i.e the energy at the end of each cycle against the number of shear cycle $N_{cycle}$ for different pin concentration $\rho_{pin}$ in 3D in Fig.\ref{fig:7},\ref{fig:8},\ref{fig:9},\ref{fig:10}. For each panel we showed two extreme temperature cases — highest T corresponds to the most poorly annealed sample and lowest T corresponds to the most well-annealed sample within the temperature range we have studied for each $\rho_{pin}$. Stroboscopic energies with cycles at different $\gamma_{max}$ at $\rho_{pin} = 0$ is plotted for the poorly annealed ($T_p = 1.0$) and well-annealed ($T_p = 0.35$) samples in Fig.\ref{fig:7}(A) and (B) respectively. We fitted the E($\gamma=0$) vs $N_{cycle}$ graph using the stretched exponential function: $E(N_{cycle}) = E_0 + b\exp\left(-\left(\frac{N_{cycle}}{\tau}\right)^{\beta}\right)$ to extract the steady-state energies and the relaxation time $\tau$ for each $\gamma_{max}$. $\tau$ represents the number of cycles to reach the steady state sheared at an strain amplitude $\gamma_{max}$. Stroboscopic energies vs $N_{cycle}$ at different $\gamma_{max}$ for the poorly annealed and well-annealed samples are also shown for $\rho_{pin} = 0.05, 0.10, 0.20$ in 3D in Fig.\ref{fig:8},\ref{fig:9},\ref{fig:10} respectively. Similarly, the plots for $\rho_{pin} = 0$ and 0.10 in 2D are also shown in Fig.\ref{fig:11},\ref{fig:12} respectively.

\section{Mode coupling temperature ($T_{MCT}$)}
\begin{figure}[H]
  \centering
   \includegraphics[width=0.32\textwidth]{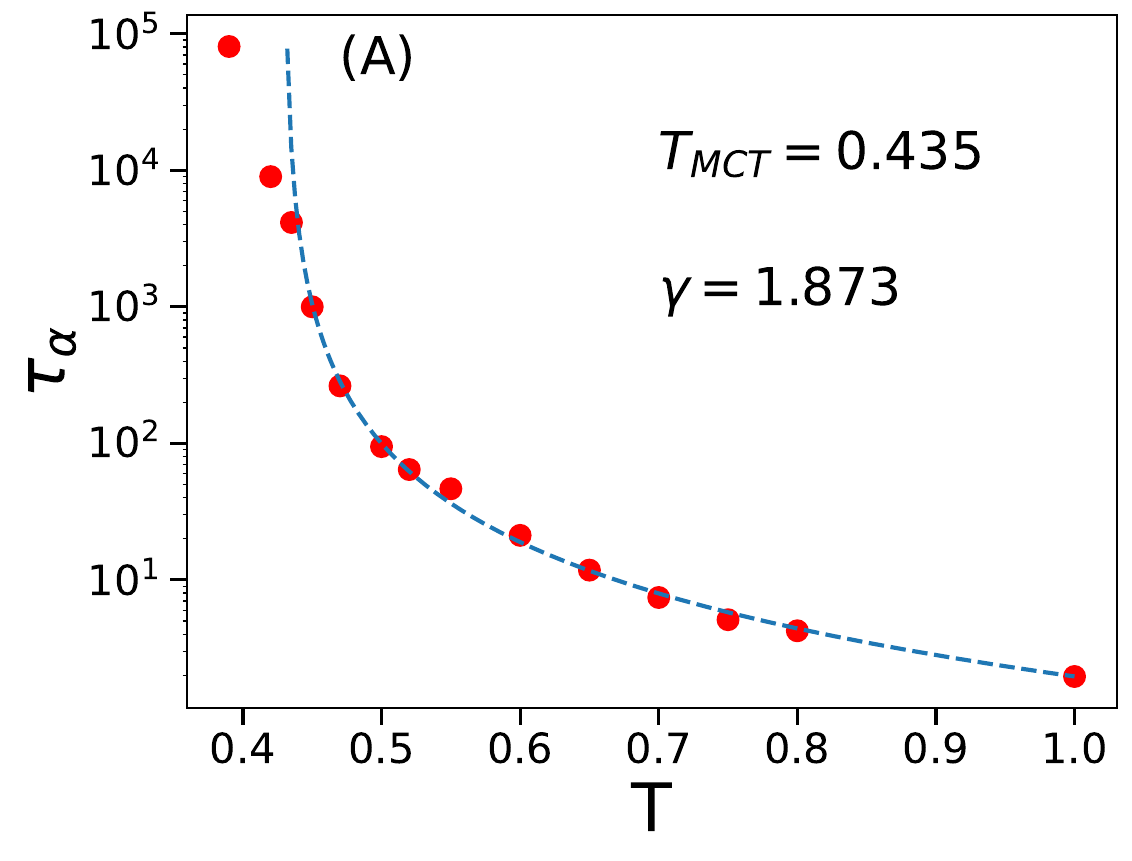}
   \includegraphics[width=0.32\textwidth]{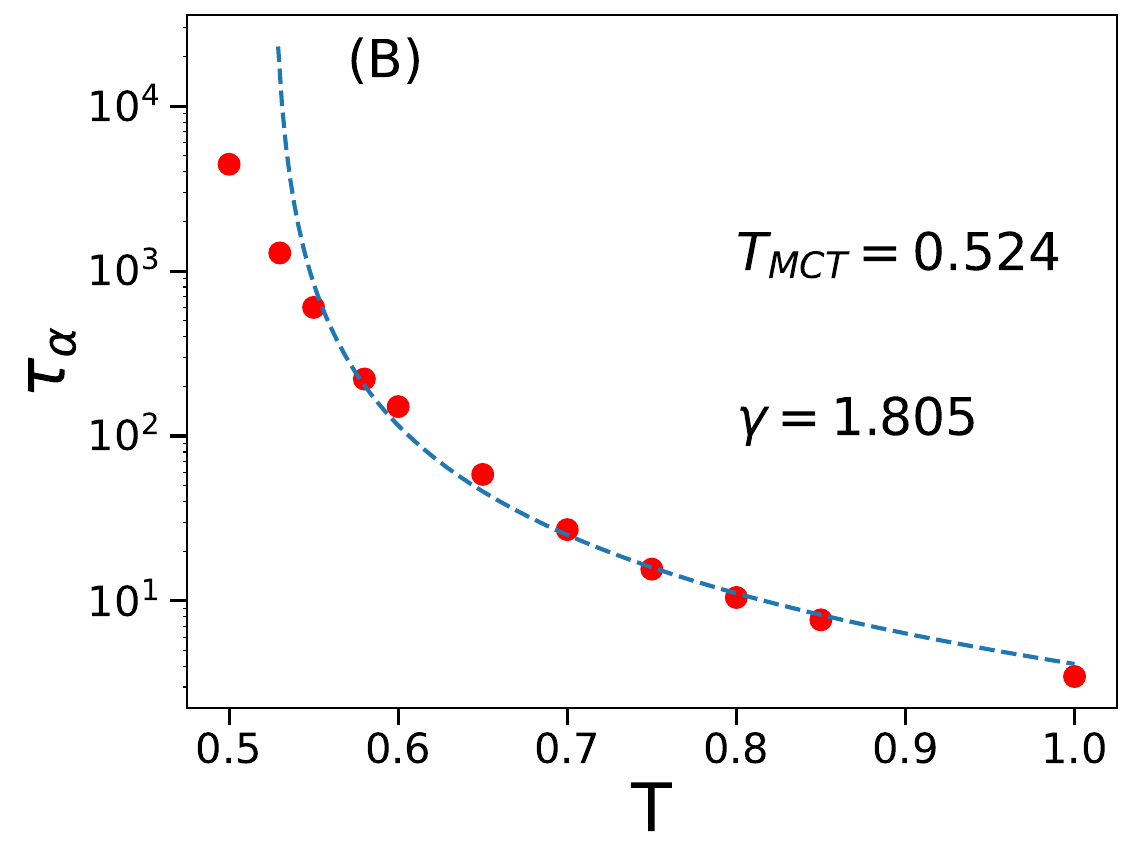}
   \includegraphics[width=0.32\textwidth]{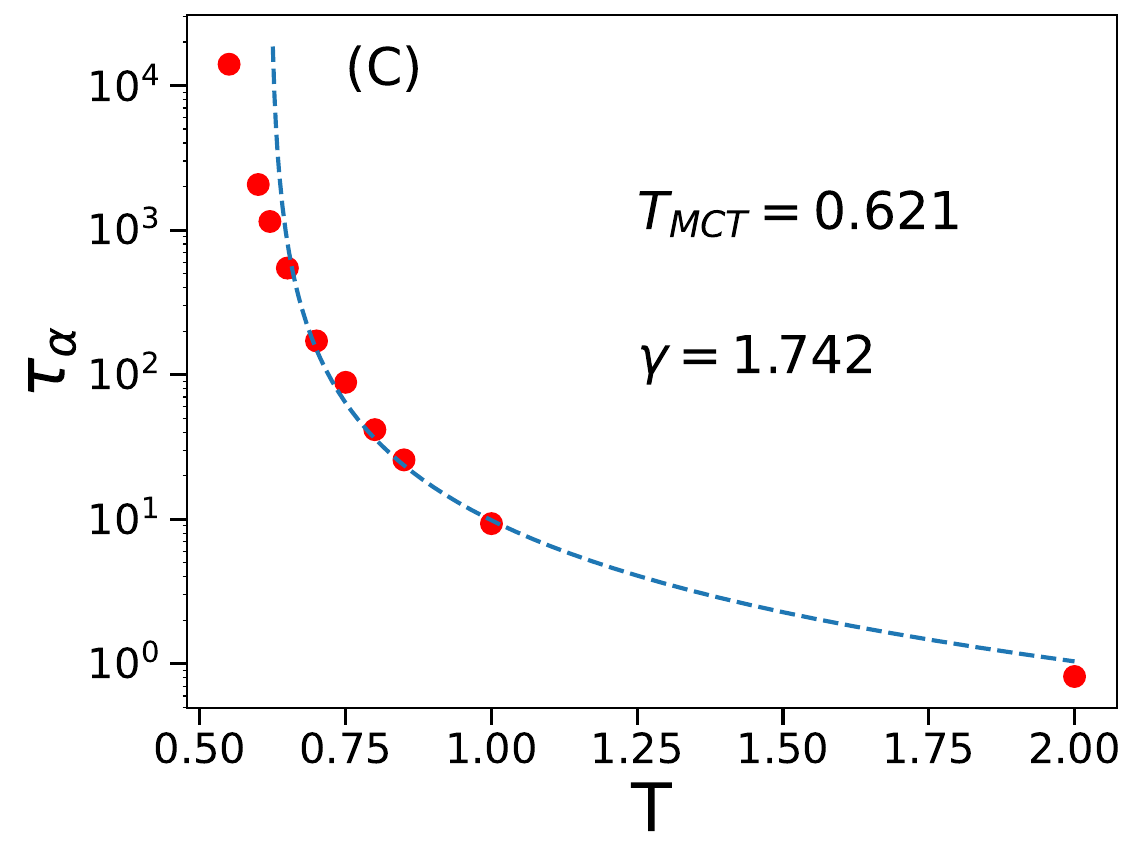}
   \includegraphics[width=0.32\textwidth]{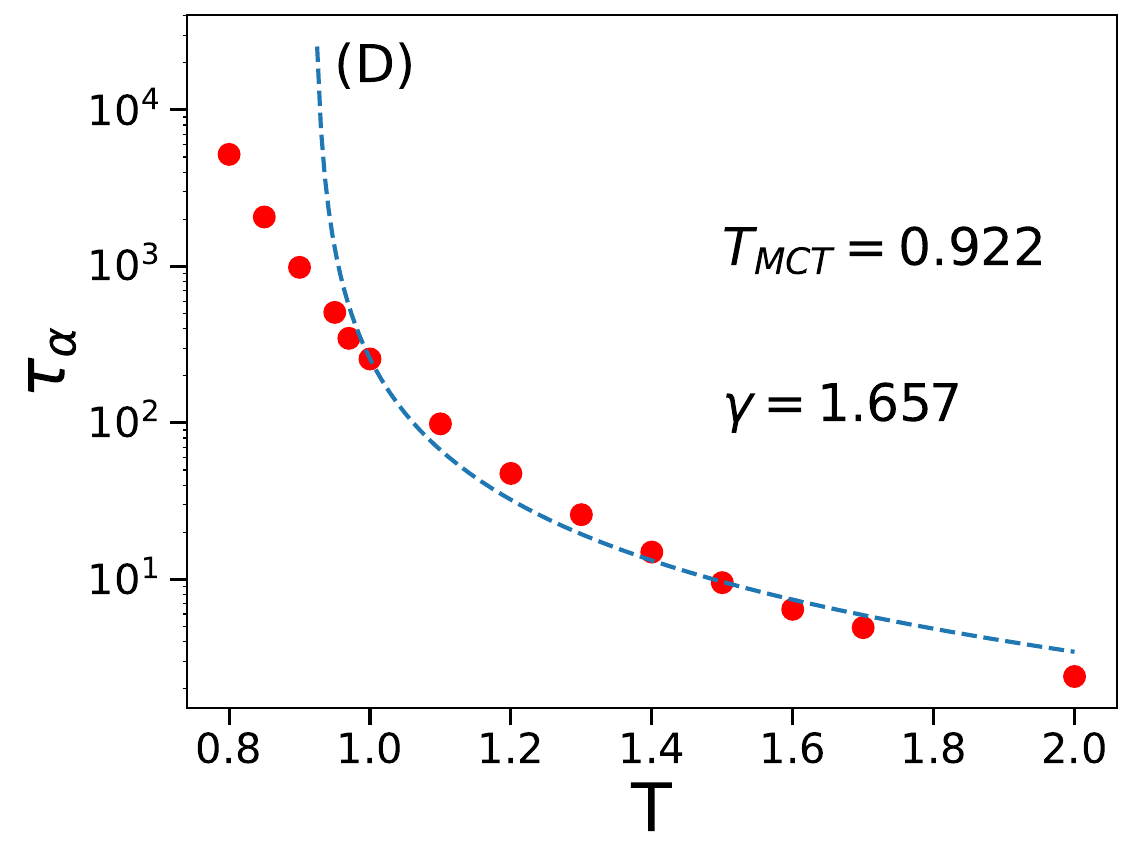}
   \includegraphics[width=0.32\textwidth]{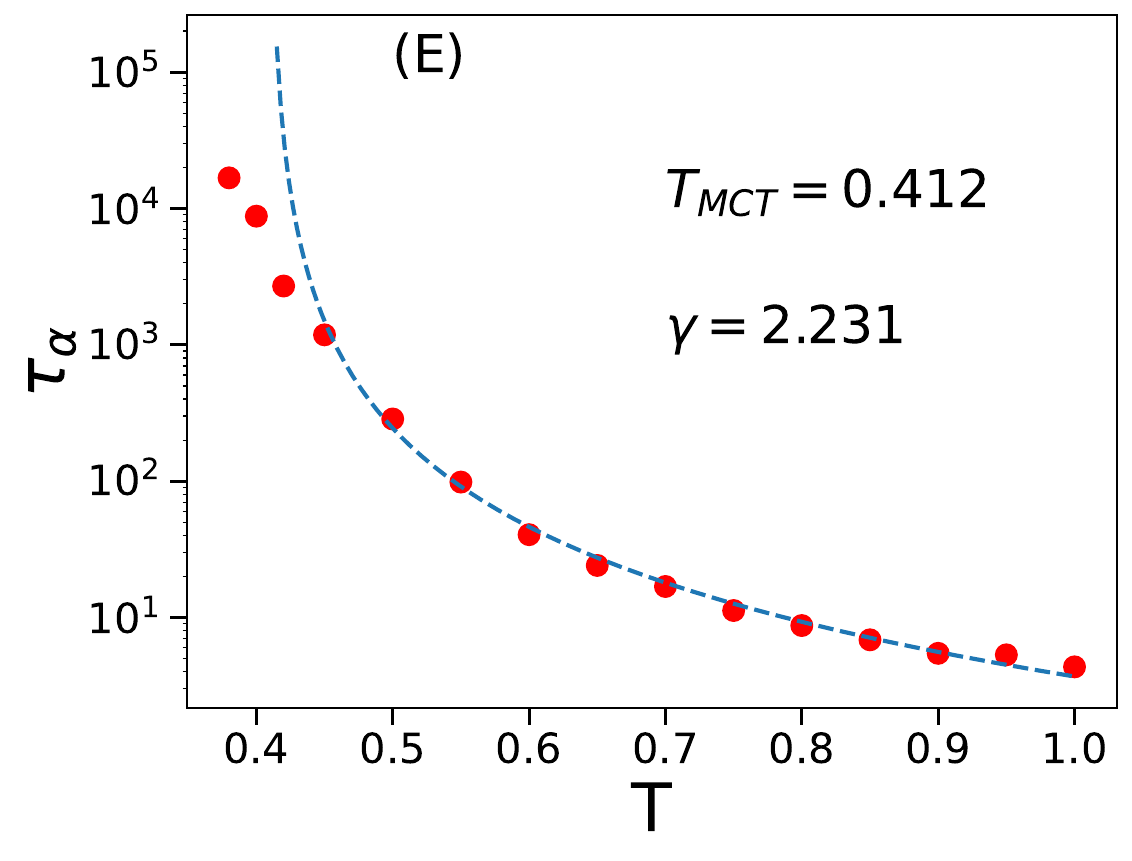}
   \includegraphics[width=0.32\textwidth]{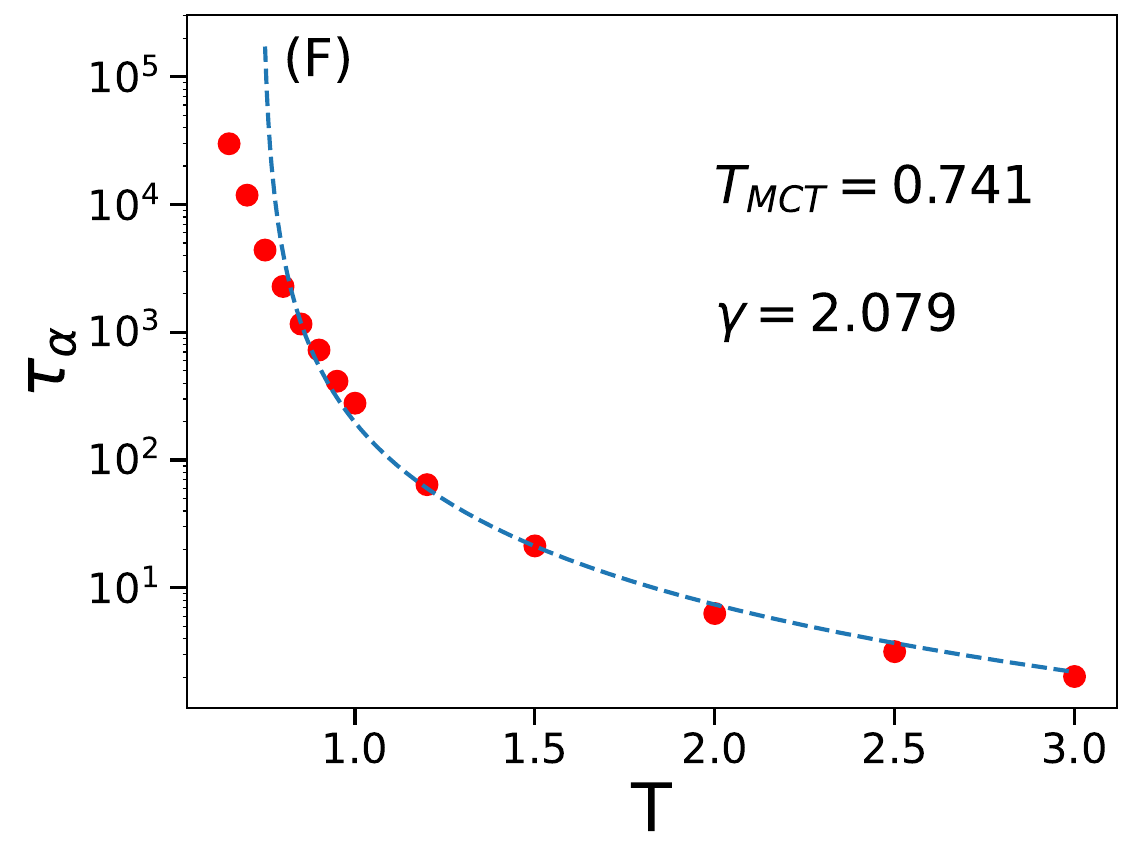}
  \caption{Plot of relaxation time $\tau_\alpha$ vs temperature T for pin concentration (A) $\rho_{pin} = 0$, (B) $\rho_{pin} = 0.05$, (C) $\rho_{pin} = 0.10$, (D) $\rho_{pin} = 0.20$ in 3D and (E) $\rho_{pin} = 0$, (F) $\rho_{pin} = 0.10$ in 2D. Data points are fitted via power law, $\tau_\alpha = a(T - T_{MCT})^{-\gamma}$.}
\label{fig:13}
\end{figure}

According to the Mode-Coupling theory, relaxation time ($\tau_\alpha$) varies with temperature (T) via a power-law form: $\tau_\alpha = \tau_0 (T - T_{MCT})^{-\gamma}$, where $\tau_\alpha$ diverges at $T = T_{MCT}$. In Fig.\ref{fig:13}, we have plotted $\tau_{\alpha}$ as a function of $T$ for different pin concentration $\rho_{pin}$ along with fits (dashed lines). $T_{MCT}$ for each case is shown within the figures.
